\documentstyle{mn}

\begin{document}
\journal{Sussex preprint SUSSEX-AST 95/11-3, Bartol preprint BA-95-52, 
astro-ph/9511057}
\title[Critical-density dark matter models]{Pursuing parameters for 
critical-density dark matter models}
\author[A. R. Liddle et al.]{Andrew R.~Liddle,$^1$ David 
H.~Lyth,$^2$ R.~K.~Schaefer,$^3$ Q.~Shafi$^3$ \cr and Pedro 
T.~P.~Viana$^1$\\
$^1$Astronomy Centre, University of Sussex, Falmer, Brighton BN1 9QH\\
$^2$School of Physics and Chemistry, University of Lancaster, Lancaster LA1 
4YB\\
$^3$Bartol Research Institute, University of Delaware, Newark, 
Delaware 19716,~~~USA}
\maketitle
\begin{abstract}
We present an extensive comparison of models of structure formation with 
observations, based on linear and quasi-linear theory. We assume a critical 
matter density, and study both cold dark matter models and cold plus hot 
dark matter models. We explore a wide range of parameters, by varying the 
fraction of hot dark matter $\Omega_{\nu}$, the Hubble parameter $h$ and the 
spectral index of density perturbations $n$, and allowing for the 
possibility of gravitational waves from inflation influencing large-angle 
microwave background anisotropies. New calculations are made of the transfer 
functions describing the linear power spectrum, with special emphasis on 
improving the accuracy on short scales where there are strong constraints. 
For assessing early object formation, the transfer functions are explicitly 
evaluated at the appropriate redshift. The observations considered are 
the four-year {\it COBE} observations of microwave background anisotropies, 
peculiar velocity flows, the galaxy correlation function, and the abundances 
of galaxy clusters, quasars and damped Lyman alpha systems. Each observation 
is interpreted in terms of the power spectrum filtered by a top-hat window 
function. We find that there remains a viable region of parameter space for 
critical-density models when all the dark matter is cold, though $h$ must be 
less than 0.5 before any fit is found and $n$ significantly below unity is 
preferred. Once a hot dark matter component is invoked, a wide parameter 
space is acceptable, including $n\simeq 1$. The allowed region is 
characterized by $\Omega_\nu \la 0.35$ and $0.60 \la n \la 1.25$, at 95 per 
cent confidence on at least one piece of data. There is no useful lower 
bound on $h$, and for curious combinations of the other parameters it is 
possible to fit the data with $h$ as high as 0.65.

\vspace*{12pt}
\noindent
{\em You are reading the e-print archive version of this paper, which due to
space restrictions doesn't have the figures. We strongly recommend you
download the complete paper, either from {\tt
http://star-www.maps.susx.ac.uk/papers/lsstru\_$\,$papers.html} (UK) or
{\tt http://www.bartol.udel.edu/$\sim$bob/papers} (US). Alternatively, e-mail
{\tt A.Liddle@sussex.ac.uk}.}
\end{abstract}

\section{Introduction}

The concept of cosmological inflation has motivated an enormous amount of
research into the formation of structure in the Universe. It has long been
known that the simplest particle physics models for inflation typically 
predict that the Universe is spatially flat and that the gravitational seeds 
for structure are adiabatic, Gaussian-distributed density fluctuations with 
a nearly Harrison--Zel'dovich spectrum (power spectrum index of $n\sim 1$).  
In order to proceed with detailed calculations from this starting point, one 
needs to pick a value for the Hubble constant. The standard choice has been 
to assume $h=0.5$, where the present Hubble constant is parametrized as 
$H_0 \equiv 100\ h$ km s$^{-1}$ Mpc$^{-1}$. Finally, the character of the 
dark matter must be decided.  Big bang nucleosynthesis implies that the bulk 
of dark matter cannot be baryonic.  

The choice for this dark matter that involves the fewest assumptions is 
relic neutrinos, as we know they exist and expect that they fill the 
Universe. Neutrinos are referred to as hot dark matter (HDM) because they 
remain relativistic until the horizon size is comparable to large-scale 
structures. Unfortunately HDM does not give a satisfactory picture of 
structure formation, because galaxies form too late and the phase space of 
the haloes of small galaxies is not large enough to accommodate the required 
number of neutrinos. A more popular alternative for the dark matter is to 
assume the existence of a cold relic particle, known as cold dark matter 
(CDM). Typical candidate particles for CDM are axions and the lightest 
supersymmetric particles.  

Initial studies of galaxy formation found that the galaxies in CDM models 
were too clustered when used with inflationary-type fluctuations (see e.g. 
Davis et al. 1985). This problem was surmounted by introducing the concept 
of {\it biasing}, in which the fluctuations in the galactic distribution are 
much larger or `biased' compared with the underlying density field. This 
model (with the ingredients $\Omega_{{\rm CDM}}=0.95$, $\Omega_{{\rm 
B}}=0.05$, $h=0.5$ and strongly biased scale-invariant adiabatic Gaussian 
fluctuations) proved to be quite successful at explaining many properties of 
galaxies and clusters, mostly on smaller scales.  However, because the 
amplitude of density fluctuations needed to be reduced to account properly 
for galaxies, this also meant that there would be insufficient power for 
making much larger scale structures.  

Standard CDM's problems with large-scale structure had already
been noticed in the 1980s.  In particular, the observed spatial correlation 
of galactic clusters was much stronger than predicted by the model.  In the
meantime, it was noted \cite{hls83,SNASA,ms83} that certain realistic
particle physics grand unification models predict the simultaneous
presence of cold and hot dark matter.  Such a mixture was recognized 
(Shafi \& Stecker 1984) to have the desirable properties of reduced 
small-scale power to make galaxies properly, while still having significant 
amounts of power on larger scales. [Models that mixed hot and warm dark 
matter were also studied (Bonometto \& Valdarnini 1984; Fang, Li \& Xiang 
1984; Valdarnini \& Bonometto 1985).] However, it was pointed out 
\cite{AOS85,BBE,vDS} that, if the mixture contained more HDM than CDM, the 
model would have difficulty forming galaxies early enough.  Thus this model 
became known as the cold plus hot dark matter model (CHDM), to indicate that 
there should be more CDM than HDM.  

As evidence for more large-scale power than expected in CDM models continued 
to accumulate during the 1980s the outlook for the CHDM model brightened.  
The number densities of cosmic structures and large cluster correlation 
lengths in CHDM models were shown to be in better agreement with 
observations \cite{OcchS88,vDS,HP}. In addition to showing that CHDM 
predictions of observed large-scale bulk flows agreed better with 
observations, two papers predicted the amplitude of the temperature 
fluctuations expected on large angular scales in the cosmic microwave 
background radiation (Schaefer, Shafi \& Stecker 1989; Holtzman 1989) well 
before the launch of the {\it Cosmic Background Explorer} ({\it COBE}) 
satellite.  The verification by {\it COBE} of these anisotropy predictions 
brought about a wave of intense interest in the CHDM model.  

The recent rapid increase in the quality of the observations of large-scale 
structure and microwave background temperature fluctuations has led to a new 
precision in investigations of theoretical models. Until recently, it was 
standard practice to derive conclusions about models of structure formation 
within a fairly rigid subset of assumptions about the relevant parameters. 
Two of these parameters are the spectral index $n$, and a parameter $r$ 
specifying the relative contribution of gravitational waves to the cosmic 
microwave background anisotropy \cite{LL92}, and they have usually been 
fixed at the canonical values $n=1$ and $r=0$. However, within a given model 
of inflation their values are determined or at least constrained, and while 
many models do accurately give these canonical values there are other models 
which do not. It is therefore realistic to allow $n$ and $r$ to vary when 
confronting a model with the data, and we shall adopt that viewpoint in this 
paper. An observational determination of these parameters in the future will 
be a powerful constraint on models of inflation, and hence on the nature of 
the fundamental interactions at very high energy scales. (For recent 
discussions of the relation between models of inflation 
and the fundamental interactions, see e.g. Schaefer \& Shafi 1994; Copeland 
et al. 1994; Dvali, Shafi \& Schaefer 1994; Stewart 1995a,b; Banks et al. 
1995; Randall, Solja\u{c}i\'{c} \& Guth 1995; Ross \& Sarkar 1996.) 

Another parameter which is often fixed is the Hubble constant, usually to 
the value $0.5$. Its variation can be important: the amount of small-scale 
power is extremely sensitive to the value of the Hubble parameter, as the 
redshift of matter domination scales quadratically with $h$. The choice of 
baryon density also can have a modest impact, as we discuss shortly. Our 
intent here is to study the CHDM model realistically, by varying the 
parameters of inflation and $h$ to see which are the most favourable values 
by testing the model against data. 

The above set of parameters is by no means complete, even within the 
limited context of inflation. For instance, inflation says nothing about 
whether or not there might be a relic cosmological constant $\Lambda$ 
contributing to the present-day spatial flatness, although it may be 
difficult to understand the magnitude of the residual $\Lambda$ within the 
philosophical context of inflation. In keeping with the original spirit of 
inflation, we set $\Lambda=0$ here. Recently it has also been emphasized 
that one can obtain genuinely open universes from inflation, albeit at 
present only with considerable tuning of parameters. Structure formation is 
apparently viable in these models \cite{RP,GRSB,LLRV}, but we shall not 
consider them here. Our assumption, therefore, is that the Universe 
possesses a critical density of matter.  

Further impetus has been delivered to the CHDM model by the observations of
neutrino oscillations from the Sun, atmospheric cosmic ray cascades, and 
possibly by the Liquid Scintillator Neutrino Detector (LSND) experiment 
\cite{Caldwell,LSND}. These observations suggest that some of the neutrino 
masses are non-zero, and that one or more neutrino species may provide a 
significant HDM density. It has even been suggested that a multiple (2 or 
3) neutrino CHDM scenario \cite{ShaS} may provide an even better fit to 
observational data (Primack et al. 1995; Pogosyan \& Starobinsky 1995b; 
Babu, Schaefer \& Shafi 1996). While promising, this remains speculative 
physics. Here we will only consider the situation of a single few eV mass 
neutrino. Indeed such a scenario can be made reasonably compatible with all 
of the oscillation experimental results (see Babu et al. 1996 and references 
therein). We are also, of course, assuming the standard cosmology for the 
neutrinos (for some alternative proposals, see Kaiser, Malaney \& Starkman 
\shortcite{KMS}; Bonometto, Caldara \& Masiero \shortcite{BCM}; Pierpaoli \& 
Bonometto \shortcite{PB}; Pierpaoli et al. \shortcite{PCBB}).

While detailed $N$-body simulations, necessitating the selection of 
particular parameter values, are required to provide a detailed comparison 
of models against observations, it is vital to carry out an investigation of 
the wider parameter space using the less intensive strategy of linear 
perturbation theory in order to find those regions of parameter space best 
suited to matching the data. Linear theory and quasi-linear theory offer 
powerful tools for investigating the shape of the density perturbation 
power spectrum across a very wide range of scales, as there are now copious 
data addressing scales large enough to still be linear today. Further, even 
the shorter scales that are non-linear today can be investigated by 
examining phenomena such as the abundance of quasars and damped Lyman alpha 
systems at moderate redshift, corresponding to times when those scales were 
still in the linear regime.

A choice is required for the baryon density, which is taken to agree with 
standard nucleosynthesis. The theory of nucleosynthesis has seen some 
developments recently, and the range advocated by Walker et al. 
\shortcite{NUCL} is now seen as too stringent, especially their upper limit. 
We choose to take a value compatible with more recent analyses by Copi, 
Schramm \& Turner (1995a,b) and Hata et al. \shortcite{Hata} which is 
$\Omega_{{\rm B}} h^2 = 0.016$. Copi et al. (1995a,b) claim that a 
plausible range of $\Omega_{{\rm B}}$, clearly intended to be thought of as 
95 per cent confidence, extends to 50 per cent in either direction around 
that value. The slightly higher value is helpful, especially in models 
without a hot component, as it helps to remove short-scale power from the 
spectrum. In CDM models, there may be further motivation to raise it further 
towards the top end of the range (e.g. White et al. 1995b); in such models 
it is easy to quantify the benefit of raising $\Omega_{{\rm B}}$ and we 
shall show how to do this later.

For our investigation, we shall therefore treat as our three main parameters 
$\Omega_{\nu}$, $n$ and $h$, and in addition allow the incorporation of a 
gravitational wave component though the space of that parameter will not be 
as extensively explored. Early investigations typically only varied 
$\Omega_{\nu}$, but were followed by treatments by Schaefer \& Shafi 
\shortcite{SS93}, Liddle \& Lyth \shortcite{LL93b} and Schaefer \& Shafi 
\shortcite{ScSh} who investigated the $\Omega_{\nu}$--$n$ plane, both with 
and without gravitational waves but concentrating only on $n<1$. Pogosyan \& 
Starobinsky \shortcite{PogStar} carried out an analogous investigation of 
the $\Omega_{\nu}$--$h$ plane, fixing $n=1$. More recently, Pogosyan \& 
Starobinsky \shortcite{PogStar2} made a study of the full 
$\Omega_{\nu}$--$n$--$h$ parameter space, concluding that $|n-1|$ should not 
exceed 0.1 for any $h$ or $\Omega_{\nu}$. Dvali et al. \shortcite{Dvalss} 
have analysed the $\Omega_{\nu}$--$n$ plane for three values of $h$ and 
found the same trends evident in Pogosyan \& Starobinsky 
\shortcite{PogStar2}, although the limits on $n$ were somewhat dependent on 
$h$ as $0.80 \la n (h/0.5)^{1/2} \la 1.15$. An analysis solely of microwave 
anisotropies on various scales applied to tilted CHDM models \cite{GMV} 
favours low $n$ values.

Our present paper is closest in spirit to the Pogosyan \& Starobinsky 
\shortcite{PogStar2} and Dvali et al. \shortcite{Dvalss} analyses, so 
it is worth indicating here where we differ.  We make an entirely new 
calculation of the transfer functions for our models, including the 
incorporation of the baryonic component (not included by them) which is 
significant especially for low $h$ values. In addition to using the more 
modern {\it COBE} normalization, we make a recalculation of constraints 
from cluster abundance, which are now more conservative.  We include a 
treatment of damped Lyman alpha system abundance, which has been seen 
as problematic for some versions of the CHDM scenario.  We shall also use 
results from the POTENT analysis of velocity fields \cite{BD89,Dekel}, but 
we include detailed modelling of the effects of cosmic variance (as included 
in Schaefer \& Shafi 1994).  Also, only $h > 0.4$ has been 
considered previously. Regardless of one's view regarding constraints from 
direct measurement, it is worth extending this to lower values to 
investigate the `volume' of favoured parameters. Further motivation for 
this arises as models with extra massless species or decaying particles can 
mimic low values of $h$ while keeping the actual $h$, as would be directly 
measured, higher (Dodelson, Gyuk \& Turner 1994; White, Gelmini \& Silk 
1995b). 

An important subset of the parameter space which we shall also explore is 
the case of pure CDM models: that is, the case $\Omega_{\nu} = 0$. These 
have seldom been studied in the context of permitting full variation of $n$, 
$h$ and the gravitational wave amplitude, and it has recently been suggested 
\cite{WSSD} that claims that all such models are ruled out may be premature. 
Our results support the assertion that there remains some viable 
parameter space for CDM models, without one having to modify the dark matter 
content or change the number of massless species.

It is useful to have a fiducial model to make comparisons with. We shall 
adopt the usual practice of taking this to be the standard CDM (SCDM) model, 
even though this is known not to be a good fit to the data. To be explicit 
about our assumptions, the parameters of this model are $n=1$, no 
gravitational waves, $h = 0.5$, $\Omega_{{\rm B}} = 0.016h^{-2} = 0.064$, 
$\Omega_{\nu} = 0$ and the amplitude of the spectrum normalized to match the 
four-year {\it COBE} observations with expected quadrupole $Q_{{\rm rms-PS}} 
= 18.0 \, \mu{\rm K}$ \cite{G96}.

The layout of the paper is as follows. In Section \ref{THEOR} we briefly 
outline the derivation of the power spectra we use to make the comparison 
with the observations, along with some discussion of our use of 
Press--Schechter theory. In Section \ref{OBSER} we shall provide a detailed 
account of the observations we have selected in order to make comparison 
with the theoretical predictions. Our procedure is not to use the power 
spectrum itself, but instead to concentrate on the spectrum filtered by a 
top-hat window, which represents the variance of fluctuations on a given 
scale. This quantity has several advantages, and in Section \ref{OBSER} we 
shall describe how we interpret our chosen observations in terms of this 
quantity. Section \ref{CONFR} will then provide the confrontation of the 
theoretical predictions with the observations.

\section{The Theoretical Input}
\label{THEOR}

\subsection{Transfer functions}

Inflation generates Gaussian density perturbations, which implies that their 
stochastic properties can be completely described by the power spectrum. In 
almost all inflationary models, the power spectrum $P(k)$ can be accurately 
parametrized across observable scales by a power-law $P(k) \propto k^n$, 
where $k$ is the comoving wavenumber (see Liddle \& Lyth 1993a and 
references therein). The choice $n = 1$ gives the scale-invariant 
Harrison--Zel'dovich spectrum, but different inflationary models predict 
different $n$, with the overall range easily encompassing all values of $n$ 
of interest for structure formation. Inflation will also generate 
long-wavelength gravitational waves which may contribute to the {\it COBE} 
signal; these will be discussed later.

We shall use a slightly different definition of the spectrum from the usual 
one, defining the spectrum of any type of perturbation $f$ as \cite{LL93a}
\begin{equation}
{\cal P}_f(k) = 4\pi \left(Lk/2\pi \right)^3 \langle |f_k|^2 \rangle \,,
\end{equation}
where $L$ is the comoving size of the periodic box introduced to allow a 
Fourier expansion of $f$ into its comoving modes $f_k$, and the angled 
brackets indicate an averaging over a small region of $k$-space to make the 
spectrum a smooth function. We have used statistical isotropy to say that 
the spectrum can only be a function of the magnitude of $k$, and not of its 
direction. The prefactor is chosen to guarantee that the mean square 
perturbation is given by
\begin{equation}
\sigma_f^2 = \int_0^\infty {\cal P}_f (k) \frac{{\rm d}k}{k} \,.
\end{equation}
Primarily we are interested in the spectrum ${\cal P}_\delta$ of the density 
contrast $\delta$, which is related to the usual $P(k)$ by ${\cal P}_\delta 
\propto k^3 P(k)$. 

The initial spectrum generated by inflation will be modified as 
the universe evolves, since the growth of density perturbations is affected 
by the properties of the matter in the universe and also the value of the 
Hubble parameter. This modification is quantified by the {\em transfer 
function} $T(k,z)$, which measures the amount of growth that a perturbation 
on scale $k$ receives by a redshift $z$ relative to the infinite wavelength 
$k = 0$ mode (thus $T(k,z) \rightarrow 1$ on large scales). With a power-law 
initial spectrum from inflation, at a given redshift one has
\begin{equation}
{\cal P}_{\delta}(k,z) \propto k^{3+n} T^2(k,z) \,,
\end{equation}
where the constant of proportionality is to be fixed via observations.

For reasons discussed in the next section, we choose not to try to place 
constraints directly on the power spectrum. Instead, we choose the 
dispersion $\sigma(R)$ of the density contrast smoothed on a scale $R$ as 
our fundamental quantity. The smoothing is carried out via a top-hat window 
function, defined by
\begin{equation}
W(kR) = 3 \left( \frac{\sin(kR)}{(kR)^3} - \frac{\cos(kR)}{(kR)^2} 
	\right) \,,
\end{equation}
which filters out modes with $k^{-1} \ll R$. The variance of the smoothed 
field is
\begin{equation}
\sigma^2(R) = \int_0^\infty W^2(kR) \, {\cal P}_{\delta}(k) \frac{{\rm
	d}k}{k} \,.
\end{equation}
Often the spectrum is increasing towards short scales, in which case the 
variance is dominated by modes with $k^{-1} \sim R$.

It is often useful to associate a mass with the top-hat filter, which one 
gets by integrating the filter over a uniform density. Assuming critical 
density, this yields
\begin{equation}
M(R) = 1.16 \times 10^{12} h^{-1} \left( \frac{R}{h^{-1} {\rm Mpc}} 
\right)^3 
{\rm M}_{\sun} \,.
\end{equation}

We have calculated the transfer functions numerically using the techniques
described by Schaefer \& de Laix \shortcite{SdL}.  The procedure can be 
summarized as follows. Starting from adiabatic initial conditions deep 
within the radiation-dominated epoch, the gauge-invariant linear evolution 
equations for each of the components are numerically integrated up to the 
present time via a Haming-type Predictor--Corrector.  We keep 1000 moments 
of the photon and relativistic neutrino distribution up until well into the 
matter-dominated epoch redshift $z=250$, at which time we set their 
amplitudes equal to zero.  At this time they have a negligible influence on 
the growth of the matter perturbations.  The massive neutrinos require 
special treatment.  In this case we expand the neutrino distribution 
function in terms of angular moments of the cosine of the angle between the 
particle momentum and the wavevector, keeping 200 angular moments. The 
massive neutrino distribution function must be integrated over momentum at 
every integration step, and this is done with 20-point Gauss--Laguerre 
integration which is accurate to better than one part in $10^6$.  The 
photons and baryons are treated using the tight coupling approximation until 
the temperature drops below $6000$ K, at which point we switch to the full 
equations for the two coupled components.  

We calculate transfer functions for $\Omega_\nu=0.0$, $0.1$, $0.2$, $0.3$, 
$0.4$, $0.5$ for $h=0.3$, $0.4$, $0.5$, $0.6$, $0.7$ using a value of the 
baryon fraction consistent with nucleosynthesis\footnote{We carried out 
these tests using the old nucleosynthesis value of Walker et al. 
\shortcite{NUCL}, before our decision to adopt the higher value 
$\Omega_{{\rm B}} h^2 = 0.016$ (Copi et al. 1995a,b) which is that used to 
obtain all results in this paper. This change does not affect our tests of 
the fitting quality.}, $\Omega_{{\rm B}} h^2 = 0.0125$. We calculate them 
for $z=0$, $3$, $3.5$, $4$. We have fit them with coefficients in a form 
somewhat similar to the Bardeen et al. \shortcite{BBKS} CDM transfer 
functions; however, the coefficients are not smooth functions of 
$\Omega_\nu$, which proved to be inconvenient for testing. We note that 
there already exists a `universal' transfer function for the CHDM models in 
universes with no baryons added \cite{PogStar2}, which we shall adapt to 
models with baryons. Pogosyan \& Starobinsky's transfer function begins with 
the Bardeen et al. \shortcite{BBKS} fit to standard CDM, 
which is 
\begin{eqnarray}
\label{tscdm}
T_{{\rm SCDM}}(q) & = & \frac{\ln \left(1+2.34q \right)}{2.34q}
	\times \\ \nonumber
  & & \hspace*{-1.7cm}
	\left[1+3.89q+(16.1q)^2+(5.46q)^3+(6.71q)^4\right]^{-1/4} \,,
\end{eqnarray}
where the scaled wavenumber $q$ is related to the usual Fourier wavenumber 
$k$ as $q=k/h^2$. Pogosyan \& Starobinsky \shortcite{PogStar2} then 
constructed a formula for the factor that describes the damping of the 
massive neutrino component:
\begin{equation}
\label{psd}
D(q,z) = \left[\frac{1 +(Aq)^2 + 
a_{{\rm eq}}(1+z)(1-\Omega_\nu)^{1/\beta}
	(Bq)^4} {1 +(Bq)^2 - (Bq)^3 + (Bq)^4}\right]^\beta \,,
\end{equation}
where 
\begin{eqnarray}
\label{pspar}
\beta &=& \frac{5}{4} (1-\sqrt{1-24\Omega_\nu/25}) \,; \nonumber \\
A &=& 17.266 \frac{ (1 + 10.912 \Omega_\nu)\sqrt{\Omega_\nu
	(1-0.9465\Omega_\nu)} }{1 + (9.259 \Omega_\nu)^2} \,; \nonumber \\
B &=& 2.6823 \frac{1.1435}{\Omega_\nu + 0.1435} \,; \nonumber \\
a_{{\rm eq}} &=& \frac{4.212\times 10^{-5}}{h^2} \,.
\end{eqnarray}

A widely used empirical formula for adding the effect of baryons to the
standard CDM transfer function is to generalize the formula for $q$ 
to $q = k/h\Gamma$ where the `shape parameter' $\Gamma$ is defined by 
$\Gamma = h \exp(-2\Omega_{{\rm B}})$. This is known to work well for CDM 
provided that $\Omega_{{\rm B}}$ is not too large \cite{PD}. We then tested 
whether or not this worked with the Pogosyan \& Starobinsky transfer 
functions.  We found that this replacement works extremely well provided 
that $h \ga 0.5$ and $\Omega_{{\rm B}} \la 0.1$. As $h\rightarrow 0.21$ the 
decoupling time approaches the time of matter--radiation equality, 
so the damping of fluctuation growth by the baryons recedes in importance.  
We have found that, for $h>0.21$, a better replacement is to use 
\begin{equation}
\label{shape}
\Gamma = h \exp \left[-2(1-(0.21/h)^2) \, \Omega_{{\rm B}} \right] \,,
\end{equation}
which is very accurate for $\Omega_{{\rm B}} \la 0.1$.  This relation 
also holds for pure CDM transfer functions. If $\Omega_{{\rm B}} \ga 0.1$, 
the baryons then become dynamically significant and impose a steep drop at 
the decoupling length scale, a feature which cannot be adequately described 
by simply shifting the scales in the transfer function.  For $h=0.3$, we 
have the central value $\Omega_{{\rm B}}= 0.139$ and notice significant 
departures of the scaled Pogosyan \& Starobinsky transfer function from our 
computed functions. We can compare values of the dispersion $\sigma(R)$ 
calculated using the real and the scaled transfer functions. In this case we 
find that the scaled functions overestimate the amplitude $\sigma(R)$ by
as much as 10 per cent on small scales $\sim 1 h^{-1}$ Mpc and  
underestimate it on scales $\sim 200 h^{-1}$ Mpc.  For comparison, when 
$h=0.4$, implying $\Omega_{{\rm B}} = 0.078$, the error in $\sigma(R)$ is 
less than about 2 per cent when $R > 0.1 h^{-1}{\rm Mpc}$. At larger values 
of the Hubble constant the fits are even better. 

Using the value $\Omega_{{\rm B}} h^2 = 0.016$ that we adopt to obtain 
results in this paper, the accuracy of the scaled Pogosyan \& Starobinsky 
transfer function becomes worse for small values of $h$; a good fit to our 
computed functions across the range of scales that we are interested in can 
only be achieved for $h > 0.4$, instead of the previous limit $h > 0.35$. 
However, for Hubble constant values as low as these we find that using the 
exact transfer functions leads to slightly stronger constraints, so adopting 
the fitting function as above is a conservative choice.

Putting all this information together, the redshift-dependent transfer 
function for CHDM models is given by
\begin{equation}
T(k,z) = T_{{\rm SCDM}}(k) D(k,z) \quad ; \quad q = k/h\Gamma \,,
\end{equation}
where $\Gamma$ is given by equation (\ref{shape}).

Some of the observations that we use apply at moderate redshift rather than 
redshift zero. In a cold dark matter dominated universe this can easily be 
accounted for using the scale-independent linear growth law $\sigma(R) 
\propto(1+z)^{-1}$, implying a redshift-independent transfer function at 
late times. By contrast, in CHDM models the growth rate becomes 
scale-dependent with suppression on short scales due to neutrino 
free-streaming. Fig.~1 illustrates the redshift dependence of the transfer 
function for two choices of HDM density. We see that, on scales greater than 
$3 h^{-1}$ Mpc, the CDM growth law is an excellent approximation from 
moderate redshift even when a sizeable HDM component is present. 

\subsection{Gravitational waves from inflation}
\label{gravwaves}

In addition to generating a power-law spectrum of density perturbations, 
inflation generates a power-law spectrum of gravitational wave modes 
\cite{Star79,LL93a}. The only observation we discuss that these are capable 
of influencing is the {\it COBE} observation, where a possible gravitational 
wave contribution to microwave background anisotropies \cite{AW,Star85} will 
add in quadrature to that from density perturbations.

Within the usual slow-roll inflation models, the amplitude of gravitational 
waves on {\it COBE} scales is another free parameter, independent of the 
spectral index of density perturbations\footnote{However, the spectral index 
of the gravitational wave spectrum is then related to the amplitude via a 
`consistency relation'.} \cite{LL92}. We shall treat the amplitude as given 
independently; the independent choice of $n$ and the gravitational wave 
amplitude is then the most general outcome of slow-roll inflation for any 
choice of potential for the scalar field driving inflation \cite{LL93b}. 

If one were to be more specific in the choice of inflation model, then the 
spectral index and gravitational wave amplitude could be related. For 
example, power-law inflation yields $n < 1$ and $r \simeq 2\pi (1-n)$, 
where $r$ is the relative contribution of gravitational waves to density 
perturbations to large-angle microwave background anisotropies\footnote{In 
some papers, the relative amplitude of gravitational waves and density 
perturbations is given as $7(1-n)$. This refers to the relative 
contributions to the quadrupole, which has a correction from the curvature 
of the last scattering surface. The version we give is appropriate to higher 
multipoles, and since the {\it COBE} normalization is most sensitive around 
the tenth multipole it is the best version to use in this context.}, as 
defined by Liddle \& Lyth \shortcite{LL92}. Almost all known inflation 
models have gravitational wave contributions sandwiched between zero and 
that of a power-law inflation model yielding the same spectral index. We 
shall concentrate on these two options for $n<1$, and ignore gravitational 
waves for $n>1$ since it is hard to make inflationary models with $n>1$ and 
significant gravitational waves.

\subsection{Press--Schechter theory}

The standard comparisons that we make between theory and observations, based 
on the spectrum integrated with a top-hat filter, are well established in 
the literature. The exception is the calculations based on object abundance, 
which contain greater theoretical uncertainties than other measures, and so 
we shall discuss in depth the way that we carry this out. The standard 
technique is Press--Schechter theory \cite{PS}, which has been compared in 
depth with $N$-body simulations (e.g. Lacey \& Cole 1993, 1994), and we 
shall use it to obtain constraints on the abundances of each of damped Lyman 
alpha systems, quasars and galaxy clusters. 

When one applies a smoothing window with a given radius to a Gaussian 
random density field, one obtains the corresponding {\it smoothed} density 
field which is also a Gaussian random field provided that its dispersion is 
smaller than one. It is then straightforward to obtain the fraction of space 
in the universe occupied by regions where the {\em linearly} evolved 
smoothed density contrast exceeds some given threshold value. The insight of 
Press \& Schechter was to assume that for the correct threshold value this 
fraction could be identified with the fraction of matter in the universe 
that is part of gravitationally bound objects with a certain minimum mass, 
the relation between the size of the regions and the minimum mass of the 
bound objects depending on the smoothing window applied to the underlying 
density field. 

A problem with this assumption is that in linear theory half the volume of 
the universe is always composed of regions with a negative smoothed density 
contrast, and therefore only half of all the matter in the universe is 
available to form bound structures, which clearly does not happen in the 
real Universe. This problem arises because one is not taking into account 
the matter in the regions whose linearly evolved density contrast does not 
exceed the threshold value, and thus are not considered to be bound 
according to the above criterion, but which are part of bigger regions whose 
linearly evolved density contrast does exceed the threshold value, and are 
therefore bound. The original Press--Schechter derivation tries to allow for 
the matter in those regions simply by assuming that they contain as much 
matter as is contained within the regions that are bound according to the 
original criterion. Though this assumption makes some sense if one thinks in 
terms of the statistics of a Gaussian random field, the main motivation was 
that it is the simplest way of allowing all the matter in the universe to be 
available to form gravitationally bound structures. This is less than 
satisfactory, and since then many people have tried in all sort of ways to 
determine if this assumption has any validity. The conclusion reached from 
$N$-body simulations is that it depends on the smoothing window, being a 
reasonable assumption for a window that is a top-hat in $k$-space, known as 
a sharp-$k$ window (for which Peacock \& Heavens \shortcite{PH} and Bond et 
al. \shortcite{BCEK} have proven using analytic methods that the factor 
two correction is exact\footnote{Recently Yano, Nagashima \& Gouda 
\shortcite{YNG} have recovered this result using a different technique, 
first proposed by Jedamzik \shortcite{J95}, which relies on the use of the 
integral equation of the mass function.}) and for the real space top-hat 
window that we use, but not so good for a Gaussian window \cite{LC}. We use 
the top-hat window as the relation between the size of a region and its mass 
is then straightforward, which is not the case for the sharp-$k$ window.

The density in collapsed objects above a given mass at a redshift $z$ is 
then given simply by integrating over the tail of the Gaussian with the 
additional factor two multiplier, yielding
\begin{equation}
\label{ps}
\Omega(>M(R),z) = {\rm erfc} \left( \frac{\delta_{{\rm c}}}{\sqrt{2} \, 
\sigma(R,z)} \right) \,,
\end{equation}
where $\delta_{{\rm c}}$ is the threshold value, $\sigma(R,z)$ is the 
dispersion smoothed on scale $R$ at redshift $z$ and `erfc' is the 
complementary error function.

The choice of threshold is crucially important, as typically it is one of 
the main sources of uncertainty. The literature features a wide range of 
values, but it is vital to note that this is primarily because different 
types of smoothing window require different thresholds. Once a specific 
choice of window is made the uncertainty is not so great. In the original 
manifestation of the Press--Schechter theory, a threshold $\delta_{{\rm c}}$ 
of 1.7 was motivated by the spherical collapse model for a top-hat 
perturbation. However, this is a highly idealized model which assumes that 
the collapsing perturbation is not under any external influence: that is, it 
does not possess shear. This should be an increasingly good assumption the 
less evolved the smoothed density field is \cite{Ber} and the less relative 
large-scale power there is. The influence and relative importance of shear 
on the time a perturbation takes to collapse depends critically on one's 
definition of collapse. If one identifies collapse of a perturbation with 
collapse along the first collapsing axis then shear decreases the time-scale 
of the collapse, but if one identifies collapse of a perturbation with 
complete collapse along three normal axes then shear increases this 
timescale \cite{Mo}. The first case relates to the formation of pancakes, 
and for example can be useful in the study of the objects that give rise to 
the Lyman alpha forest lines in the spectra of quasars. However, if one is 
interested in completely virialized objects like quasars or clusters than 
the second definition of collapse should be used. The damped Lyman alpha 
systems are likely to lie somewhere in between these two extremes. 

To a large extent, the analytic modeling of the threshold has been 
superseded by direct calibration of the Press--Schechter theory with 
$N$-body simulations. Indeed, if one were to take an extreme view one could 
regard the Press--Schechter formula simply as a fitting function to the 
number density at a given epoch. Comparison with $N$-body simulations 
indicates that for the top-hat filter the spherical collapse estimate 
$\delta_{{\rm c}} = 1.7$ actually works extremely well for virialized 
objects, with at most an uncertainty of $0.2$ in either direction \cite{LC}. 

It is often emphasized that the predicted number density can depend very 
sensitively on the choice of threshold and on the dispersion, especially 
where the dispersion is small. This is of great advantage for this 
application, because it means that, even if there is a large observational 
uncertainty in the number density, this gives only a small uncertainty in 
the estimate of the dispersion. Concerning the uncertainty in the threshold, 
we can see directly from the Press--Schechter formula above that an 
uncertainty of 12 per cent in $\delta_{{\rm c}}$ translates into the same 
uncertainty in the estimate of $\sigma(R)$.

\section{The Observational Data}
\label{OBSER}

To constrain the density perturbation spectrum effectively, one requires a 
compilation of estimates of its amplitude at a variety of different scales 
by a variety of different methods. To some extent, this occurs naturally as 
different types of observations are best suited to estimating the power 
spectrum on different scales. For example, only microwave background data 
are presently capable of providing information on the largest scales, and 
only the abundance of objects at high redshift allows access to presently 
non-linear scales at a time when they may still be addressed using 
quasi-linear theory. It is only the intermediate scales, running from 
perhaps $8 h^{-1}$ Mpc up to $100 h^{-1}$ Mpc, that have been simultaneously 
constrained by a number of different types of measurements, from abundance 
of clusters to galaxy correlation functions to peculiar velocity flows; in 
the near future reliable microwave background experiments should also extend 
down into this region. 

Our general strategy is not to impose constraints on the power spectra 
$P(k)$ directly (where $k$ is the comoving wavenumber), but instead to 
impose them on the dispersion of the density field filtered through a 
top-hat window function, denoted $\sigma(R)$, whose radius $R$ is varied in 
order to pick out different scales. This method is useful because the bulk 
of the observational data are obtained in this form, and typically the 
conversion of such data into power spectrum form introduces systematic 
errors. In contrast, a theoretical calculation of the filtered variance is 
very simple to make given a theoretical power spectrum. Further, 
observations on short scales connected with object formation at high 
redshift have no interpretation at all in terms of the power spectrum at a 
given wavenumber; the standard method of theoretical comparison using the 
Press--Schechter calculation deals directly with the filtered variance. 
Concentrating on estimating a single function such as this has the advantage 
that to a large extent the data can be presented together and treated on the 
same footing.

\subsection{{\it COBE}}

Recently there has been considerable activity concerning the interpretation 
of the anisotropies detected by {\it COBE} \cite{COBE,B94,W94}. The 
four-year data set is now available \cite{Ben96,Ban96,G96,H96} and we 
will use results from it in this paper. The methods used have become 
sufficiently sophisticated that simple normalization methods, such as to the 
$10\degr$ variance of the anisotropies as widely used in the two years 
following the {\it COBE} announcement, are no longer appropriate, since they 
make inadequate use of the full {\it COBE} data set. Instead, it is better 
to rely on the normalizations published in the literature which do take the 
full data set into account. 

The first development in this regard was a very elegant pair of papers by 
G\'{o}rski and collaborators \cite{Gorski,Getal} who fitted power-law 
spectra to the observations, thus obtaining likelihoods in the $n$--$Q_{{\rm 
rms-PS}}$ plane, where $Q_{{\rm rms-PS}}$ is the expected quadrupole (over 
an ensemble of independent observers). The quantity that is actually desired 
in order to constrain theoretical models is not the full likelihood or the 
marginalized one, but rather the conditional likelihood on $Q_{{\rm 
rms-PS}}$ for fixed $n$, given as a function of $n$. Although they only 
provided this for $n = 1$, they noted that regardless of the fitted $n$ the 
preferred amplitude of the ninth multipole remains unchanged to excellent 
accuracy, and this result can be used to generate the required normalization 
as a function of $n$. 

However, more recently it has been noted that the assumption of a 
power-law spectrum of anisotropies, corresponding to the Sachs--Wolfe 
contribution, is not a perfect one, because the `Doppler peak' extends to 
small multipoles and invades part of the region that {\it COBE} samples. 
Consequently, one should fit the amplitude using full anisotropy 
spectra. This was carried out for the two-year {\it COBE} data by Bunn, 
Scott \& White \shortcite{BSW}. For CDM spectra with no gravitational waves, 
they find conditional likelihoods yielding
\begin{equation}
Q_{{\rm rms-PS}}(n) = (19.9 \pm 1.5) \exp \left[ 0.69(1-n) \right] \; 
	\mu {\rm K} \quad \mbox{(2yr)} \,.
\end{equation}
Note that the $n$ dependence in the fit given was calculated only taking 
into account the Sachs-Wolfe effect; it should nevertheless provide a very 
good approximation. Bunn et al. (1995) noted that this result is more or 
less independent of the nature of any dark matter, of $\Omega_{{\rm B}}$ and 
of $h$, so it can be used for all models without gravitational waves. 
Although the full anisotropy spectra are needed for performing the fit to 
the {\it COBE} data, it is fine to compute the perturbation spectrum 
normalization corresponding to a given quadrupole using the Sachs--Wolfe 
formula, since the quadrupole is least affected by the Doppler peak.

We need to correct this for the new four-year data, for which amplitudes 
conditional on $n$ have not yet been published. However, the principal 
change, due largely to a new galactic cut strategy, is a lowering of the 
normalization without changing the shape information. It is therefore fine 
to use the same $n$-dependence with the lowered normalization (G\'{o}rski et 
al. 1996; M. White, private communication), yielding
\begin{equation}
Q_{{\rm rms-PS}}(n) = (18.0 \pm 1.4) \exp \left[ 0.69(1-n) \right] \; 
	\mu {\rm K} \quad \mbox{(4yr)} \,.
\end{equation}
This is the normalization that we shall adopt. The quoted error is 
1$\sigma$.

As we are concentrating on interpreting data in terms of the filtered 
dispersion $\sigma(R)$, it is interesting to ask what sort of scales this 
normalization is sampling. One way to do this is to normalize a set of CDM 
models with different $n$ and see where the curves cross. One finds that the 
lines more or less cross (with an accuracy of a few per cent, doing less 
well as $h$ is varied) at a scale of $4000 h^{-1}$ Mpc. We shall 
occasionally use this to represent the {\it COBE} data schematically, with 
the main purpose of indicating the size of the {\it COBE} error; however, in 
all cases we shall calculate using the precise normalization of the power 
spectrum given above rather than this approximate data point.

To be completely accurate, if gravitational waves are included one should 
add their radiation power spectrum to that of the density perturbations and 
perform a full model fit. However, as long as the gravitational wave 
contribution is not too significant, one can approximate it as having the 
same functional form as the density perturbations over the {\it COBE} range 
and simply normalize down the density perturbation power spectrum as 
appropriate. As discussed by Liddle \& Lyth \shortcite{LL93b}, the relative 
contribution of gravitational waves to density perturbations to the 
microwave anisotropies, $r$ as defined in Subsection \ref{gravwaves}, leads 
to a reduction in the amplitude of the {\it COBE} normalized dispersion 
$\sigma(R)$ by a factor $1/\sqrt{1+r}$.

Although we will normalize $\sigma(R)$ to the G\'{o}rski et al. 
\shortcite{G96} {\it COBE} central value, we shall allow for their 15 per 
cent uncertainty at the 2$\sigma$ level by adding it in quadrature to the 
relative errors of those other observations which also constrain the 
amplitude of $\sigma(R)$.

\subsection{Galaxy correlations}

The deficiencies in the shape of the standard CDM spectrum are most apparent 
in surveys of galaxy correlations spanning the range from a few megaparsecs 
up to tens of megaparsecs. A variety of surveys such as QDOT, CfA, APM and 
1.2 Jansky provide information in this region. In an attempt to evade 
systematics particular to the types of analysis provided, one can combine 
data from a variety of different sources, hoping to demonstrate consistency 
between the different data sets, and this has been achieved in an 
impressive analysis by Peacock \& Dodds \shortcite{PD}. The cost is that the 
formal errors are somewhat larger than those one sees in individual surveys, 
and it is not easy to see whether or not one is unfairly penalizing the most 
accurate data sets rather than uncovering overoptimistically small error 
bars across all data sets.

Another problem with using galaxy data is that, although they determine the 
shape of the spectrum very well, the overall normalization is less certain 
due to the expectation that galaxy correlations are biased relative to the 
underlying matter, multiplying the power spectrum by a (hopefully 
scale-independent at least over the limited range of scales considered) bias 
parameter. One can attempt to determine the bias parameter from the surveys 
themselves by using redshift distortions and/or non-linear effects, or 
instead by utilizing an entirely separate method such as peculiar velocity 
flows. Alternatively one can allow the normalization of the galaxy 
correlation data to `float', with its best amplitude determined by the other 
types of data under consideration, which amounts to throwing away 
information on the bias.

Peacock \& Dodds \shortcite{PD} quote their final results in terms of the 
power spectrum ${\cal P}_{\delta}(k)$ ($\Delta^2(k)$ in their notation). 
However, the original data are provided in a mixture of the power spectrum, 
the dispersion $\sigma(R)$ and the correlation function $\xi(R)$. They 
switch between them using an analytic prescription:
\begin{eqnarray}
\sigma(R) &=& {\cal P}_{\delta}^{1/2}(k_R) \,; \\
\xi(R) &=& {\cal P}_{\delta}^{1/2}(\sqrt 2 k_R) \,,
\end{eqnarray}
where 
\begin{equation}
k_R = \left[\frac{1}{2} \Gamma \left( \frac{m+3}{2}\right)\right]^{1/
	(m+3)} \frac{\sqrt5}{R}  \,,
\end{equation}
and $m\equiv (k/{\cal P}_{\delta})({\rm d} {\cal P}_{\delta}/{\rm d} k)$ is 
the effective spectral index. These formulae are obtained by assuming $m$ 
constant over the range of $k$ modes contributing, and using the 
approximation 
\begin{equation}
W(kR)=\exp(-k^2R^2/10) \,,
\end{equation}
which is exact for $kR\ll1$.

In making the conversion, one needs to specify a power spectrum in order to 
calculate the effective spectral index. This is best done by choosing a 
model spectrum that fits the data; we use the best-fitting CDM spectrum, 
specified by a shape parameter $\Gamma$. Since the raw data are provided in 
a variety of forms, there is no reason to think that expressing them in 
terms of $\sigma(R)$ is any less accurate than expressing them via the power 
spectrum, and we shall use both. The shape parameter provides a good 
indication of the quality of fit to the galaxy correlation data and we shall 
occasionally use that language.

Recently some doubt has been cast over the assumption by Peacock \& Dodds 
that the bias parameter is scale-independent down to the smallest scales, 
around $4 h^{-1}$ Mpc, considered in their analysis \cite{P96}. As it 
seems that it is for scales below around $8 h^{-1}$ Mpc that the bias 
parameter starts becoming non-linear, we have excluded from their final 
data, presented in table 1 of Peacock \& Dodds \shortcite{PD}, the four 
points corresponding to the smallest scales and re-calculated the best 
$\Gamma$ fit to their remaining data. For $0.7<n<1.2$ we find $\Gamma = 0.23 
- 0.28(1-1/n)$, where the 2$\sigma$ relative error is +18 per cent and -15 
per cent. This compares with the central value (for $n=1$) from the full 
data set of $0.255$ \cite{PD}.

When utilizing data of this form in a statistical analysis, as we do below, 
it is vital to ensure that the points used are taken suitably far apart as 
to be independent, and in general one needs the full correlation matrix to 
determine this (which has been calculated only for the QDOT survey power 
spectrum \cite{FKP}). If one is not careful as regards this point, then 
statistical tests are biased and, depending on the form of test used, this 
can make bad models look good or, much more seriously, make good models look 
bad. For a statistical treatment there is the further problem that the 
errors are systematic as well as statistical, and hence will not be normally 
distributed; unfortunately in the absence of a detailed understanding of an 
experiment there is no way to counter this other than to treat results with 
mild scepticism. 

After the exclusion of the four points corresponding to the smallest scales, 
a chi-squared analysis of the remaining data in table 1 of Peacock \& Dodds 
\shortcite{PD}, where $n$, $h$, $\Omega_{\nu}$ and the normalization are the 
fitting parameters, has 8 degrees of freedom. Performing this analysis we 
find a very low minimum chi-squared of around 4; although it is perfectly 
reasonable that this occurred by chance, it may also indicate weak residual 
correlations of neighbouring data points. In the present case this typically 
makes models seem much better in relation to the data than they really are. 
The best way that we found of avoiding this problem is to calculate not the 
absolute exclusion level of each model against the data, but the relative 
confidence limits in the three-dimensional space formed by the parameters 
$n$, $h$ and $\Omega_{\nu}$ \cite{PTVF}. This is achieved by calculating the 
difference between the chi-square obtained for each model characterized by a 
fixed set of values for $n$, $h$ and $\Omega_{\nu}$, where the normalization 
is calculated so as to minimize the chi-squared, and the minimum chi-squared 
obtained by varying the four parameters. This difference still has a 
chi-squared distribution, now with three degrees of freedom. The 68 per cent 
and 95 per cent confidence limits are then defined in the 
$(n,h,\Omega_{\nu})$ space by the chi-squared difference being respectively 
smaller than 3.508 and 7.815. We will plot cross-sections of the region in 
the $(n,h,\Omega_{\nu})$ space defined by the 95 per cent confidence limit.

\subsection{Peculiar velocities}

\subsubsection{POTENT}

Peculiar velocities directly sample the matter power spectrum and so are 
unaffected by clustering bias. However, measurements of the peculiar 
velocity field are much harder to obtain. The best measurements using 
velocities alone come from the POTENT method \cite{BD89}, the most recent 
version available being the Mark III POTENT data \cite{Dekel}, which 
supply an estimate of the velocity smoothed on various length scales around 
us. This can be used as an estimator for $\sigma(R)$ on a particular scale, 
as follows.

First, we restrict ourselves to using a single piece of data, the velocity 
on a $40 h^{-1}$ Mpc sphere. Although measurements exist for a range of 
scales, they are very highly correlated because the window function for the 
peculiar velocities samples a wide range of scales and in particular is more 
sensitive to longer scales than the density dispersion. We choose this 
particular value as it is in the centre of the supplied range.

In making a theoretical comparison, one needs a two-stage smoothing, since 
POTENT involves first smoothing the observed peculiar velocities with a $12 
h^{-1}$ Gaussian before the velocity reconstruction can be undertaken and 
the $40 h^{-1}$ top-hat smoothing applied to obtain $v(40h^{-1} {\rm Mpc})$. 
The appropriate formula for the dispersion of the velocity is
\begin{equation}
\sigma_v^2(R) = H_0^2 \int_0^{\infty} W^2(kR) \exp \left( -(12h^{-1}k)^2
	\right) \, \frac{{\cal P}_{\delta}}{k^2} \, \frac{{\rm d}k}{k} \,.
\end{equation}

As with {\it COBE} above, one can then ask what scales in the filtered 
dispersion $\sigma(R)$ of the density field correspond to a fixed observed 
velocity. This can again be addressed by plotting $\sigma(R)$ for a set of 
CDM models with different $n$, each normalized to yield the same 
$\sigma_v(40h^{-1} {\rm Mpc})$. It turns out that such curves cross, 
extremely accurately, at a scale of $113 h^{-1}$ Mpc. As stated above, the 
velocities sample considerably longer scales than the smoothing length by 
itself suggests. 

This crossing point remains quite accurate even if one goes to CHDM models, 
and this fact coupled with the much larger observational errors as compared 
with {\it COBE} means that we can represent the POTENT data as a single 
constraint on $\sigma(113 h^{-1} {\rm Mpc})$.

The Mark III POTENT analysis gives for the bulk flow in a $40 h^{-1}$ Mpc 
sphere \cite{Dekel}
\begin{equation}
\label{bf}
v_{{\rm POTENT}}(40h^{-1} {\rm Mpc}) = 373 \pm 50 \; {\rm km \, s}^{-1}\,, 
\end{equation}
where the error arises from different ways of dealing with sampling-gradient 
bias and can thus be thought of as reflecting the systematic uncertainty in 
the POTENT analysis. Additionally there is an intrinsic uncertainty in the 
POTENT calculation due to random distance errors, which at the 1$\sigma$ 
level is $\simeq 15$ per cent \cite{Dekel}. Note that {\it COBE} normalized 
standard CDM yields
\begin{equation}
v_{{\rm SCDM}}(40h^{-1} {\rm Mpc}) = 409 \; {\rm km \, s}^{-1}\,,
\end{equation}
suggesting that SCDM produces about the right answer using the modern {\it 
COBE} normalization. The observational error is dominated by cosmic 
variance, resulting from the POTENT observation being a single measurement 
from a random field. Since each velocity component separately has a Gaussian 
distribution, the velocity squared has a chi-squared distribution with three 
degrees of freedom. From this one can calculate the range of theoretical 
values for which the observed value would not lie in the tail of the 
distribution, and the probabilities corresponding to 68 per cent confidence 
yield an upward error of $89$ per cent and a downward error of $24$ per cent 
on the estimator for $\sigma(R)$. At the $95$ per cent confidence level the 
error bars are $+273$ per cent and $-43$ per cent. The asymmetry of the 
errors originates in the asymmetry of the chi-squared distribution. We can 
now convolve the systematic and random errors arising from the POTENT 
calculation with the cosmic variance error. Assuming that the error in 
expression (\ref{bf}) corresponds to something like $95$ per cent confidence 
(though as it is the smallest error this assumption is insignificant), we 
then obtain the total error in using the Mark III POTENT bulk flow 
calculation as an estimator of the normalization of the dispersion of the 
density contrast: at the $68$ per cent confidence level, $+98$ per cent and 
$-25$ per cent; at the $95$ per cent confidence level, $+295$ per cent and 
$-47$ per cent. Clearly, only the lower limits are of use for us. At a level 
corresponding to 95 per cent confidence, the bulk flow constraint can then 
be written as 
\begin{equation}
\frac{\sigma_{{\rm POTENT}}(113 h^{-1} {\rm Mpc})}{\sigma_{{\rm SCDM}}
	(113 h^{-1} {\rm Mpc})} = 0.91_{-47 {\rm \; per \; cent}}^{+295 
	{\rm \; per \; cent}}\,.
\end{equation}

\subsubsection{Velocities versus densities}

An alternative use of velocity data is through the comparison with the 
density field obtained via galaxy surveys. Present technology focuses on an 
estimate of a single parameter $\Omega_0^{0.6}/b$, where $b$ is the bias 
parameter appropriate to whatever type of galaxies is being studied, 
normally {\it IRAS} galaxies with bias $b_I$. The degenerate combination of 
$\Omega_0$ and $b$ arises through the inability to distinguish slow 
velocities due to a slowing of the perturbation growth rate in low-density 
universes from having a high irregularity in the galaxy distribution 
relative to that of the matter distribution actually generating the 
velocities. However, we are considering only critical-density models, so 
these methods directly estimate the bias. This information can then be used 
in conjunction with the galaxy number counts dispersion to supply 
constraints on the variance in the density. Note though that there seems no 
good way to quantify the errors arising from the inadequacy of a single bias 
parameter to explain the difference between the galaxy and density 
variances.

We shall not utilize the range of bias found by Peacock \& Dodds 
\shortcite{PD}, the reason being that there remains widespread disagreement 
in the literature between values obtained by different methods (for 
instance, see Dekel 1994 for a compilation). Consequently, the true 
uncertainty appears much greater than advertized by any single study, and if 
one attempts to take a more realistic view the amplitude becomes so 
uncertain as to provide no useful constraint.

\subsection{Abundance of galaxy clusters}

The typical mass of large galaxy clusters, about $10^{15} \, {\rm 
M}_{\sun}$, corresponds to a linear scale of around $8 h^{-1}$ Mpc. 
Observation indicates that large clusters are relatively rare, suggesting 
that this scale is still in the quasi-linear regime. The usual technique of 
Press--Schechter theory calibrated by $N$-body simulations can therefore be 
used to impose constraints. A variety of estimates of the cluster mass 
function exist in the literature; some authors (e.g. Lilje 1992; White, 
Efstathiou \& Frenk 1993a) aim to reproduce the observed number density at a 
single mass scale whilst others (e.g. Evrard 1989; Henry \& Arnaud 1991; 
Hattori \& Matsuzawa 1995) more ambitiously aim to fit the shape of the 
cluster mass function. The analysis we perform belongs to the first type. 
Typically, the number density of a given type of cluster is quite well 
known, at least at low redshift --- most of the uncertainty comes from poor 
knowledge of the mass of individual clusters. This can be estimated in a 
variety of ways, the most common being the virial theorem, the X-ray 
temperature distribution as a tracer of the gravitational potential and, 
most recently, weak shear lensing of background objects. All these methods 
suffer from several problems, though the one that at the present seems most 
likely to give the best results is the use of X-ray temperature 
observations. 

The observed number density of clusters per unit temperature at $z=0$ about 
a mean X-ray temperature of 7 keV was calculated by Henry \& Arnaud 
\shortcite{HA} to be 
\begin{equation}
\label{nd}
n(7 \; {\rm keV} , 0) = 2.0_{-1.0}^{+2.0}\times10^{-7}h^{3} 
	\; {\rm Mpc}^{-3} \; {\rm keV}^{-1} \,.
\end{equation}

The comoving number density of clusters with virial mass $M_{{\rm v}}$ per 
mass interval ${\rm d}M_{{\rm v}}$ at a redshift $z$ is obtained by 
differentiating equation (\ref{ps}) with respect to the mass and multiplying 
it by $\rho_{{\rm b}}/M_{{\rm v}}$, where $\rho_{{\rm b}}$ is the comoving 
background density (a constant during matter domination), thus giving
\begin{eqnarray}
\label{mfa}
n(M_{{\rm v}},z) \, {\rm d}M_{{\rm v}} = & & \nonumber \\
& & \hspace*{-2cm} -\sqrt{\frac{2}{\pi}} \frac{\rho_{{\rm b}}}{M_{{\rm v}}}
	\frac{\delta_{{\rm c}}}{\Delta^2(z)} \frac{{\rm 
d}\Delta(z)}{{\rm
	d}M_{{\rm v}}} \exp \left[-\frac{\delta_{{\rm 
c}}^2}{2\Delta^2(z)}
	\right] {\rm d}M_{{\rm v}} \,,
\end{eqnarray}
where $\Delta\equiv\sigma(r_{{\rm L}})$ with $r_{{\rm L}}$ the comoving 
linear scale associated with $M_{{\rm v}}$, $r_{{\rm L}}^3=3M_{{\rm 
v}}/4\pi\rho_{{\rm b}}$. Traditionally the cluster abundance is used to 
constrain the present-day dispersion at $8h^{-1}$ Mpc, $\sigma_{8} \equiv 
\sigma(8h^{-1} {\rm Mpc}, 0)$, and the quantity $\Delta$ is specified by an 
analytic approximation to the power spectrum in the vicinity of this scale. 
Generally, one can write
\begin{equation}
\label{del}
\Delta(z)=\sigma_{8}(z) \left(\frac{r_{{\rm L}}}{8h^{-1} \; 
	{\rm Mpc}}\right)^{-\gamma(r_{{\rm L}})} \,.
\end{equation}
In Liddle et al. \shortcite{LLRV} we adopted the form
\begin{equation}
\gamma(r_{{\rm L}}) = (0.3\Gamma+0.2) \left[2.92 + \log 
	\left(\frac{r_{{\rm L}}}{8 h^{-1} \; {\rm Mpc}}\right)\right]\,, 
\end{equation}
where $\Gamma$ is a shape parameter. Though this fit is strictly only 
correct for scale-invariant pure CDM models, it can also be used as a 
fitting function for the dispersion of the observed linear power spectrum on 
some restricted range of scales, which for our purposes means within a 
factor of 1.5 of 8$h^{-1}$ Mpc. The values used for $\Gamma$ will then be 
those allowed by observations, i.e. $\Gamma \in [0.19,0.27]$ at the 
2$\sigma$ confidence level\footnote{Using the Peacock \& Dodds 
\shortcite{PD} 2$\sigma$ interval, $\Gamma\in[0.22,0.29]$, does not change 
the final results.}.

Note that, unlike with pure CDM models, the shape of the power spectrum for 
CHDM models is not redshift independent since the growth of perturbations at 
a given scale depends on the mean random peculiar velocities of the massive 
neutrinos at the scale in question which in turn are redshift dependent. 
However, for the scales of interest for clusters, in the CHDM models that we 
consider the redshift evolution of the shape of the power spectrum is 
extremely small in the redshift interval where most clusters form in these 
models, i.e. $z \leq 0.5$. 

Using expression (\ref{del}) to calculate the derivative in equation 
(\ref{mfa}), we therefore get
\begin{eqnarray}
\label{mf}
n(M_{{\rm v}},z) \, {\rm d}M_{{\rm v}} = & & \\ \nonumber
& & \hspace*{-2cm} \sqrt{\frac{2}{\pi}}
	\frac{\rho_{{\rm b}}}{M_{{\rm v}}^{2}}
	\frac{2.92(0.3\Gamma+0.2)\delta_{{\rm c}}}{3\Delta(z)} \exp
	\left[-\frac{\delta_{{\rm c}}^{2}}{2\Delta^{2}(z)}\right] {\rm
	d}M_{{\rm v}} \,.
\end{eqnarray}

As we are considering clusters massive enough that at the corresponding 
scale the density field is not yet well developed into the non-linear 
regime, according to the discussion on Subsection 2.3 we can therefore 
ignore the influence of shear on their formation and assume that they 
collapsed spherically. Nevertheless, to be conservative we shall include an 
assumed 1$\sigma$ dispersion of $\pm 0.1$ in the value of $\delta_{{\rm 
c}}$, i.e. $\delta_{{\rm c}}=1.7\pm0.1$.

Using self-similar evolution arguments (e.g. Hanami 1993), which have been 
shown to be in good agreement with hydrodynamical $N$-body simulations 
(Navarro, Frenk \& White 1995), one obtains the following relation between 
the cluster virial mass, $M_{{\rm v}}$, its mean X-ray temperature, $k_{{\rm 
B}}T$, and its redshift of virialization, $z_{{\rm c}}$: 
\begin{equation}
\label{mvprop}
M_{{\rm v}} \propto (1+z_{{\rm c}})^{-3/2}(k_{{\rm B}}T)^{3/2} \,.
\end{equation}

We begin by considering the case of a CDM universe. In order to normalize 
equation (\ref{mvprop}) we use results from the hydrodynamical $N$-body 
simulations for an $\Omega_{0}=1.0$ CDM model performed by White et al. 
\shortcite{WNEF}. From a catalogue of 12 simulated clusters with a wide 
range of X-ray temperatures they estimated that a cluster with a present 
mean X-ray temperature of 7.5 keV corresponds to a mass within one Abell 
radius (1.5 $h^{-1}$ Mpc) of the cluster centre of $M_{{\rm A}}=(1.10 \pm 
0.22)\times10^{15} \, h^{-1} \; {\rm M}_{\sun}$. The error arises from the 
dispersion in the catalogue and is supposed to represent the 1$\sigma$ 
significance level. White et al. \shortcite{WNEF} also found that the 
simulated clusters had a density profile in their outer regions 
approximately described by ${\rho}_{{\rm c}}(r)\,\propto\,r^{-2.4\pm0.1}$. 
This same result was obtained by Metzler \& Evrard \shortcite{ME} and 
Navarro et al. \shortcite{NFW}. Bearing in mind that the cluster virial 
radius in a $\Omega_{0}=1.0$ universe encloses a density 178 times the 
background density, it is then straightforward to calculate the cluster 
virial mass from $M_{{\rm A}}$. Through a Monte Carlo procedure, where we 
assume the errors in $M_{{\rm A}}$ and in the exponent of $\rho_{{\rm 
c}}(r)$ to be normally distributed, we find $M_{{\rm v}} = (1.23 \pm 0.32) 
\times 10^{15} \, h^{-1} \; {\rm M}_{\sun}$ for a cluster with a present 
mean X-ray temperature of 7.5 keV in an $\Omega_{0}=1.0$ universe. Assuming 
that such a cluster virialized at a redshift of $z_{{\rm c}} \simeq 0.05 \pm 
0.05$ (e.g. Metzler \& Evrard 1994; Navarro et al. 1995), we can now 
normalize equation (\ref{mvprop}): 
\begin{eqnarray}
M_{{\rm v}} = (1.32 \pm 0.34) \times 10^{15} \times & & \\ \nonumber
 & & \hspace*{-2cm} (1+z_{{\rm c}})^{-3/2} \left(\frac{k_{{\rm B}}T}{7.5\,
 	{\rm keV}} \right)^{3/2} \, h^{-1} \; {\rm M}_{\sun}\,.
\end{eqnarray}
This result is in very close agreement with the one obtained by Evrard 
\shortcite{E} from his own hydrodynamical $N$-body simulations. Hence the 
virial mass $M_{{\rm v}}$ for a cluster with a present mean X-ray 
temperature of 7 keV is given by  
\begin{equation}
\label{mv}
M_{{\rm v}} = (1.2 \pm 0.3) \times10^{15} \, (1+z_{{\rm c}})^{-3/2} \, 
	h^{-1} \; {\rm M}_{\sun}\,.
\end{equation}

Through some simple physical arguments, Sasaki \shortcite{Sa} used 
Press--Schechter theory to obtain an expression for the comoving number 
density of clusters per mass interval ${\rm d}M_{{\rm v}}$ about virial mass 
$M_{{\rm v}}$, which virialize in an interval ${\rm d}z$ about some redshift 
$z$ and survive until the present: 
\begin{eqnarray}
\label{ncom}
N(M_{{\rm v}},z) \, {\rm d}M_{{\rm v}} \, {\rm d}z = & & \\ \nonumber
& & \hspace*{-2cm} \left[-\frac{\delta_{{\rm c}}^{2}}{\Delta^{2}(z)}
	\frac{n(M_{{\rm v}},z)}{\sigma_{8}(z)} \frac{{\rm
	d}\sigma_{8}(z)}{{\rm d}z}\right] \frac{\sigma_{8}(z)}{\sigma_{8}} 
	\, {\rm d}M_{{\rm v}} \, {\rm d}z \,,
\end{eqnarray}
where for the type of models we are presently considering we have 
\begin{equation}
\sigma_{8}(z)=\sigma_{8}(1+z)^{-1}\,.
\end{equation}
In equation (\ref{ncom}) the expression within the square brackets gives the 
formation rate of clusters with virial mass $M_{{\rm v}}$ at redshift $z$, 
whereas the fraction outside gives the probability of these clusters 
surviving until the present. If one now assumes that at each redshift $z$ 
the cluster virial mass $M_{{\rm v}}$ in equation (\ref{ncom}) is determined 
by expression (\ref{mv}) with $z_{{\rm c}}=z$, then equation (\ref{ncom}) 
gives the comoving number density of clusters {\it per unit mass} that 
virialize at each redshift $z$ and survive up to the present such that they 
have a mean X-ray temperature of 7 keV at the present. Through the chain 
rule we can then determine the comoving number density of clusters {\it per 
unit temperature} that virialize at each redshift $z$ and survive up to the 
present such that they have a mean X-ray temperature of 7 keV at the 
present:
\begin{eqnarray}
N(k_{{\rm B}}T,z) \, {\rm d}(k_{{\rm B}}T) \, {\rm d}z & =  &  
	\frac{{\rm d}M_{{\rm v}}}{{\rm d}(k_{{\rm B}}T)}
	N(M_{{\rm v}},z) \, {\rm d}(k_{{\rm B}}T) \, {\rm d}z \nonumber \\ 
 & = & \frac{3}{2} \frac{M_{{\rm v}}}{k_{{\rm
	B}}T}N(M_{{\rm v}},z) \, {\rm d}(k_{{\rm B}}T) \, {\rm d}z \,,
\end{eqnarray}
where the second equality uses equation (\ref{mvprop}).
We therefore have 
\begin{eqnarray}
\label{nt}
N(k_{{\rm B}}T,z) \, {\rm d}(k_{{\rm B}}T) \, {\rm d}z =  & & \\ \nonumber
 & & \hspace*{-1cm} \frac{3}{2}	\frac{M_{{\rm v}}}{k_{{\rm B}}T}
	\frac{\delta_{{\rm c}}^{2}}{\Delta^{2}(z)}
	\frac{n(M_{{\rm v}},z)}{(1+z)^{2}} \, {\rm d}(k_{{\rm B}}T) \, 
	{\rm d}z \,.
\end{eqnarray}

Numerically integrating this expression from $z=0$ to $z=\infty$ then gives 
the present comoving number density of clusters per unit temperature with a 
mean X-ray temperature of 7 keV as a function of the present value of 
$\sigma_{8}$. Comparing with the observational value given 
by equation (\ref{nd}) we then find to a good approximation that 
\begin{equation}
\label{final}
\sigma_{8}=0.60_{-0.15}^{+0.19} \,.
\end{equation}
The errors in equation (\ref{final}) represent 95 per cent confidence levels 
and arise from the dispersions in the observational value of $\Gamma$, in 
the assumed value for $\delta_{{\rm c}}$, and in expressions (\ref{nd}) and 
(\ref{mv}). They were estimated via a Monte Carlo procedure, the full 
details of which are given by Viana \& Liddle \shortcite{VL}. That paper 
also demonstrates that essentially the same constraint can be obtained using 
the more detailed merging picture due to Lacey \& Cole (1993, 1994).

However, this result applies only to models where all the dark matter is 
cold. We would now like to know how this result is affected if one 
substitutes part of the cold dark matter by massive neutrinos. 

In galaxy clusters the X-ray emission comes mainly from a nearly isothermal 
core, and thus strongly depends on the depth and width of its gravitational 
potential. Outside the core the shape of the gravitational potential is of 
much less importance to the total X-ray emission. It is then possible to 
have galaxy clusters with the same mean X-ray temperature at virialization 
but slightly different virial masses. Though for a given cosmological model 
this dispersion should be quite small, the differences in virial mass 
between galaxy clusters with the same mean X-ray temperature at 
virialization in two different cosmological models could be significantly 
higher. 

Whilst the dependence of cluster density profiles on the slope of the power 
spectrum at the cluster scale has been studied quite thoroughly \cite{CER}, 
the consequences of changing the nature of some of the dark matter have not 
been so extensively studied. In the case of interest to us, where only one 
neutrino species has a cosmologically significant mass, the typical cluster 
density profile has been determined only for a pure neutrino model \cite{C} 
and for a model with $\Omega_{\nu} = 0.3$ \cite{KKPH}. As the fraction of 
massive neutrinos is increased at the expense of the same amount of cold 
dark matter, the depth and width of the gravitational potential at the 
nearly isothermal core, and therefore the core mass and radius, should 
remain approximately the same for clusters with equal mean X-ray temperature 
at virialization. However, we now have a component that clusters less, 
therefore leading to a more extended mass distribution. For a power-law 
density profile $\rho\propto r^{-\alpha}$, this corresponds to a smaller 
$\alpha$. It can then easily be shown that the cluster virial mass 
increases. This increase will be greater either if more CDM is substituted 
by HDM or if the neutrino free-streaming length is increased by making them 
lighter\footnote{If the neutrino mass is so small that the neutrinos are 
unable to cluster at the scales we are considering, around 2$h^{-1}$ Mpc, 
the cluster virial mass will not increase as the neutrinos will not be 
gravitationally bound to the cluster. However, the effect on the 
relationship between $\sigma_{8}$ and the cluster number density will be 
exactly the same as if the cluster virial mass had increased in reality, as 
will become clear in the section dealing with damped Lyman alpha systems.}. 
In reality these two effects oppose each other as the neutrino mass 
increases with $\Omega_{\nu}$. In the limit where all the dark matter is 
composed of massive neutrinos, these are sufficiently massive that, at the 
scales corresponding to high-mass clusters, the clustering behaviour of the 
massive neutrinos seems to resemble closely that of cold dark matter 
\cite{C}. If $\Omega_{\nu}$ is between 0 and 1, then for high-mass galaxy 
clusters we have very little information about the clustering properties of 
massive neutrinos on the scales in which we are interested, and therefore 
about the virial masses one should expect in such models. 

To our knowledge there is only one hydrodynamical $N$-body simulation study 
\cite{Bryan} that has tried to relate $\sigma_{8}$ to the abundance of X-ray 
clusters for a CHDM model. Though their resolution is insufficient to 
determine the internal density distribution of the galaxy clusters that they 
obtain, we can use the Press--Schechter approximation to re-normalize their 
simulation. First we need to calculate the present-day cluster virial mass 
$M_{{\rm v}}$ which corresponds to a present mean 
X-ray temperature of 7 keV by using their normalization of the power 
spectrum, $\sigma_{8}=0.606$, and the cluster number densities they obtain 
for that X-ray temperature. Using expression (\ref{nt}) and assuming 
$\delta_{{\rm c}} = 1.7 \pm 0.1$, through a Monte Carlo procedure as before 
we get $M_{{\rm v}}=(1.30_{-0.29}^{+0.36}) \times 10^{15} h^{-1} \; {\rm 
M}_{\sun}$ at the 1$\sigma$ confidence level for a 
mean X-ray temperature of 7 keV, where we have read the cluster number 
density from Fig.~1 of Bryan et al. \shortcite{Bryan} to be $n(7 \; {\rm 
keV} , 0) = (1.6_{-0.8}^{+1.6}) \times 10^{-7} h^{3} \; {\rm Mpc}^{-3} \; 
{\rm keV}^{-1}$. We can now use the calculated $M_{{\rm v}}$ to obtain the 
normalization that corresponds to the observed abundance of present-day 
galaxy clusters with mean X-ray temperature of 7 keV, which is given by 
equation (\ref{nd}). Again using $\delta_{{\rm c}} = 1.7 \pm 0.1$ and a 
Monte Carlo procedure, we obtain $\sigma_{8}=0.62_{-0.14}^{+0.17}$. The 
errors represent 95 per cent confidence limits, and hence both the central 
value and the size of the uncertainty are very similar to those that we got 
for a pure CDM model, where $\sigma_{8} = 0.60_{-0.15}^{+0.19}$ at 95 per 
cent confidence. It is encouraging that two rather different calculations 
give such similar answers. We shall use the relative errors obtained for a 
pure CDM model as they are slightly more conservative, and model the shift 
in the central value due to a change in $\Omega_{\nu}$ by a simple linear 
fit:
\begin{equation}
\sigma_8 = \left(0.60 + 0.2 \Omega_{\nu}/3 \right)_{-24 {\rm \; per 
	\; cent}}^{+32 {\rm \; per \; cent}} \,,
\end{equation}
where the uncertainty is 95 per cent confidence. This relation will hold 
well for the models in which we are interested (indeed, it would be 
satisfactory just to employ the CDM result and ignore the slight shift in 
central value brought on by the hot component).

\subsection{Abundance of high-redshift objects}

To constrain the present-day power spectrum on scales around $1h^{-1}$ Mpc 
requires detailed numerical simulations as those scales are well into the 
non-linear regime. However, a convenient alternative exists in the abundance 
of objects at high redshifts, which can sample the spectrum on those scales 
while they were still in the quasi-linear regime. The constraints on the 
{\em linear} power spectrum can then be evolved to the present day. In this 
context it is vital to recall that, when a hot dark matter component is 
introduced, perturbations on these scales can have their growth affected, 
typically growing more slowly than in a CDM model which has the effect of 
making the constraints weaker than na\"{\i}ve expectations. It is common to 
use analytic treatments based on rather nebulously defined neutrino Jeans 
masses to make this correction. Although this is often fine (since the 
corrections are typically small), we shall instead use direct calculations 
of the transfer functions at the appropriate redshift.

The most important objects for our purpose are quasars and damped Lyman 
alpha systems, and we shall place particular emphasis on the latter as they 
provide stronger constraints. Uncertainties as to the efficiency of quasar 
formation and the number of quasar generations mean that only a lower bound 
on the power spectrum can be obtained from them at present. Damped Lyman 
alpha systems, on the other hand, in principle also offer an upper limit 
(though to our knowledge one has never been quoted), and indeed the 
evolution of the amount of gas in such systems as a function of redshift may 
well imply significant constraints on star formation.

For each object type, it is important to be as conservative as possible in 
supplying limits; the standard strategy is to obtain a rigid constraint that 
all models are compelled to satisfy, rather than a number with an error bar 
which can be subjected to a statistical test.

\subsubsection{Quasars}

The type of power spectra we are considering flatten towards short scales, 
so that, when one studies short scales, the relative influence of 
perturbations from larger scales becomes more important. Thus, in accordance 
with the discussion in Section 2.3, we should expect shear to become more 
important, and therefore the relative time efficiency for the formation of 
bound objects to decrease, as one considers the formation of increasingly 
smaller objects. Bearing this in mind, one should then expect the 
virialized dark haloes associated with the formation of galaxies to assemble 
more slowly than those for clusters due to the presence of a relatively 
stronger shear field\footnote{Some studies \cite{AC} suggest that the 
presence of substructure within a collapsing object can increase its 
time-scale of collapse through dynamical friction. Though this effect, 
similarly to shear, delays collapse, its dependence on the shape of the 
power spectrum is the opposite, thus effectively diminishing the overall 
dependence of the time-scale of collapse on it. However, this effect seems 
not to be nearly as important as shear \cite{Mo}.}. Even if these 
suppositions turn out to be correct it is difficult to quantify precisely 
both the strength of the shear field for a given scale at a certain epoch 
and its relation with the value one should consider for $\delta_{{\rm c}}$. 
It is due to this limitation that we will use in our analysis the abundance 
of the most luminous quasars at a redshift of $z=4$, when the density field 
at the scale associated with the virialized dark haloes in which this type 
of quasar is embedded, which we will assume to have masses in excess of 
$10^{12} h^{-1} \; {\rm M}_{\sun}$, is still not well developed and 
consequently shear can be ignored to a good approximation. We can therefore 
assume that these dark haloes collapsed nearly spherically and accordingly 
use the $\delta_{{\rm c}}$ associated with spherical collapse. We will also 
assume that the time lag between halo virialization and quasar ignition is 
negligible. Adopting the most conservative result given by Haehnelt 
\shortcite{H93} as corresponding to 95 per cent confidence, we have
\begin{equation}
\sigma(M = 10^{12} h^{-1} \; {\rm M}_{\sun},z=4) \geq 0.26 \,,
\end{equation}
for an assumed quasar number density of around $5 \times 10^{-8} \; 
h^{3} \; {\rm M}_{\sun} \; {\rm Mpc}^{-3}$. The corresponding comoving scale 
is $R = 0.95 h^{-1}$ Mpc. However, this constraint is always weaker than 
that coming from damped Lyman alpha systems.

\subsubsection{Damped Lyman alpha systems}

At low and intermediate redshifts, $z\leq2$, the most popular view is 
that the vast majority of the damped Lyman alpha lines which appear in the 
spectra of quasars are produced by neutral hydrogen present in quiescent 
large-scale discs, similar to those presently found in spiral galaxies. 
However, these disc systems would have to be 2 to 3 times bigger in size 
than present spiral galaxies in order to explain the apparent increase in 
filling factor with redshift, if one assumes that the comoving number 
density of these systems remains constant (Lanzetta, Wolfe \& Turnshek 
1995). An alternative explanation would be that this increase in filling 
factor with redshift is instead due at least partially to an increase in the 
comoving number density of disc systems with redshift, in particular for $1< 
z <2$. The excess number of systems would then disappear by merging, 
possibly giving rise to some of the presently observed elliptical galaxies. 
At higher redshifts, $z\geq2$, there are some hints that these lines may be 
produced in objects more akin to turbulent protospheroids, from the apparent 
short time-scales of consumption of the neutral gas by star formation and 
the discrepancy between the observed low metallicities associated with the 
lines at those redshifts and the expected higher metallicity of the gas if 
it is to be the material from which disc stars in present spiral galaxies 
formed \cite{LWT}. These protospheroids are the natural progenitors of 
galaxies, and the indication would then be that the transition between 
turbulent collapsing haloes and quiescent rotationally supported discs 
occurred at $z \sim 2$. 

Instead of the widely quoted data of Lanzetta et al. \shortcite{LWT}, we use 
the more recent data of Storrie-Lombardi et al. \shortcite{storrie} which 
revise downwards\footnote{Note that this still ignores the effect of 
gravitational lensing, which it is claimed can reduce the estimated 
abundance by a further 50 per cent \cite{bartelmann}.} the estimated 
abundances at a redshift of around 3 and provide a new estimate at redshift 
4. The strongest constraint comes from the redshift 4 data, though it is not 
significantly weakened if the redshift 3 data are used instead. We will 
present constraints from both.

Following the discussion in the previous subsubsection, we are interested 
in the amount of matter associated with damped Lyman alpha systems at  
redshifts 3 and 4. As we have seen, at these redshifts the systems 
are probably collapsing protospheroids massive enough to give rise to 
rotationally supported gaseous discs. The minimum total mass needed in order 
for that to happen seems to be around $10^{10} h^{-1} \; {\rm M}_{\sun}$ 
\cite{H95}, which corresponds to a circular velocity of 77 km \,${\rm 
s}^{-1}$. It is not clear how far these systems have collapsed 
gravitationally. A reasonable, and for our purposes conservative, hypothesis 
is that they have just collapsed along the first two collapsing axes, i.e. 
`filament' formation, though the baryonic fraction of the collapsing 
material would have collapsed further through radiative cooling (e.g. Katz 
et al. 1994). Numerical studies indicate that a value of $\delta_{{\rm c}}$ 
around 1.5 is associated with the time-scale of gravitational collapse along 
the first two collapsing axes \cite{Mo}, and accordingly we shall use it in 
the Press--Schechter calculation. This gives a more conservative bound than 
$\delta_{{\rm c}} = 1.7$.

In Storrie-Lombardi et al. \shortcite{storrie}, the fraction of the critical 
density in the form of neutral gas associated with damped Lyman alpha 
systems at redshifts 3 and 4 is observed to be
\begin{equation}
\Omega_{{\rm gas}}(z=3)= (0.0017 \pm 0.0003) \, h^{-1}\,,
\end{equation}
and 
\begin{equation}
\Omega_{{\rm gas}}(z=4)= (0.0011 \pm 0.0002) \, h^{-1}\,.
\end{equation}
The total amount of matter that was involved in the formation of damped 
Lyman alpha systems at these redshifts as a fraction of the critical 
density is 
then given by 
\begin{equation}
\Omega_{{\rm DLAS}}(z)=\frac{\Omega_{{\rm gas}}(z)}{f_{{\rm
	gas}} \, \Omega_{{\rm B}}}\,,
\end{equation}
where $f_{{\rm gas}}$ is the neutral fraction of the gas in those systems, 
which conservatively we will assume to be 1, and $\Omega_{{\rm B}} = 0.016 
\, h^{-2}$ is the cosmological baryon density given by standard 
nucleosynthesis. We now have to be careful in deciding which is the 
characteristic comoving mass scale associated with these systems. If one 
takes $10^{10}h^{-1} \; {\rm M}_{\sun}$ to be the minimum mass of damped 
Lyman alpha systems then, because we do not expect massive neutrinos within 
the mass range we are considering to be able to cluster on this mass scale 
at $z\geq2$, the characteristic comoving mass scale involved in the 
formation of these systems is given by $M = 10^{10}(1-\Omega_{\nu})^{-1} 
h^{-1} \; {\rm M}_{\sun}$. That is, it is originally perturbations on this 
larger mass scale that begin to collapse, but at some point during the 
collapse of the perturbations there will be a segregation between the 
massive neutrinos and the cold dark matter, the former remaining in an 
oscillatory mode roughly at the scale of segregation (approximately equal to 
the neutrino Jeans scale) and the latter collapsing further, eventually 
leading to the formation of $10^{10} h^{-1} \; {\rm M}_{\sun}$ virialized 
objects. 

All this therefore implies that the fraction $f(>M,z)$ of the total mass 
that is involved in the formation of damped Lyman alpha systems at 
redshifts 3 and 4 is given by 
\begin{equation}
f(>M,z=3) > (0.106 \pm 0.033) \, h \,,
\end{equation}
and 
\begin{equation}
f(>M,z=4) > (0.069 \pm 0.021) \, h \,,
\end{equation}
where $M=10^{10}(1-\Omega_{\nu})^{-1}h^{-1} \; {\rm M}_{\sun}$. A 25 per 
cent uncertainty in the baryon fraction, corresponding loosely to 1$\sigma$, 
has been added in quadrature to the observational uncertainty. Since we want 
a lower bound on the density perturbation we take the 2$\sigma$ lower end of 
the error bar. Using equation (\ref{ps}), we then have to a good 
approximation the 95 per cent confidence limits
\begin{eqnarray}
\label{lim1}
\sigma(R,z=3) & > & 0.54 + 0.2 h \,; \\
\label{lim2}
\sigma(R,z=4) & > & 0.50 + 0.2 h \,,
\end{eqnarray}
for $0.3 < h < 0.7$, where $R = 0.2 \left(1-\Omega_{\nu} \right)^{-1/3} 
h^{-1}$ Mpc. In fact, the constraint is quite insensitive to the confidence 
limit chosen. We shall use the redshift 4 point as it provides the stronger 
constraint; although numerically the constraint is similar, it applies at a 
higher redshift. 

\subsection{Compilation}

Fig.~2 shows all the data we have discussed, plotted at the present epoch. 
The data on short scales, which are obtained at moderate redshift, are 
scaled to the present epoch assuming a pure CDM model (though in later 
analysis we shall shall directly apply the high-redshift transfer function). 
The {\it COBE} point is represented schematically as discussed. We have 
plotted the Peacock \& Dodds points assuming a bias parameter (for {\it 
IRAS} galaxies) of 1.1, which is the best fit for $\Omega_0 = 1$; the errors 
shown on the individual points correspond to the errors on their relative 
location, and the uncertainty in bias, $\pm 0.2$ (not illustrated in this 
Figure), then allows the entire data set to be shifted up or down.

This figure shows that the data follow a more or less continuous path, 
across a range of roughly four orders of magnitude both in linear scale and 
in the size of the dispersion. However, this large range makes the 
individual error bars very small, and were one to attempt to plot 
theoretical predictions on this figure it would be very hard to discern 
which were the best fit to the data.

In order to overcome this, we can use the knowledge that the standard CDM 
model, while unable to fit the observational data in detail, is certainly 
able to fit all of them to within a factor two or so. Consequently, we can 
greatly improve the graphical representation by plotting the observational 
data divided by the prediction of the {\it COBE} normalized standard CDM 
model. The choice of this particular model as the fiducial one is governed 
by history; it does not indicate any preference for this model over any 
other but rather is simply a graphical convenience. The data normalized to 
the standard CDM model are shown in Fig.~3.

As anticipated, the data all lie within a factor two or so of this canonical 
model, with the short-scale data falling below the prediction of {\it 
COBE}-normalized standard CDM. The Peacock \& Dodds data 
\shortcite{PD}\footnote{We have left out from this figure the two points 
corresponding to the largest scales in order to obtain a clearer picture of 
the observations as these points are very close to the POTENT point.}  are 
represented by a band, and the error bars on the end indicate the overall 
normalization uncertainty. 

When we make comparisons of theory and observations, one of the aspects we 
have to take into account is that the data are available at different 
redshifts. When one has a hot dark matter component, the growth of 
perturbations on short scales is slower than in a CDM model, and this must 
be taken into account. Rather than impose an analytic approximation to the 
different growth rate, we directly use transfer functions calculated at the 
appropriate redshift of around $z = 3.5$.

It is fortunate that the data available at the present day are on scales 
large enough that the growth rate is the same as in the CDM model for these 
moderate redshifts, as seen in Fig.~1. This means that one can shift these 
data back to a redshift 3.5 in a model-independent way\footnote{This is true 
only for CHDM models, and would not hold for open or cosmological constant 
models.}. Consequently, the simplest approach is to consider all the data as 
given at redshift 3.5. Had we plotted this, it would look exactly as Fig.~2, 
but with the vertical axis divided by a factor $4.5$ in accordance with the 
CDM growth law $\sigma(R) \propto 1/(1+z)$. Fig.~3 remains exactly the same, 
and when interpreted at this redshift one doesn't have to worry about 
perturbation growth suppression corrections in models with an HDM component.

Fig.~4 shows this data with some sample theoretical curves overlaid.

\section{Confrontation}
\label{CONFR}

Our three primary parameters are $n$, $h$ and $\Omega_\nu$. Let us first 
specialize our discussion to varying single parameters of the standard CDM 
model. Although there is no clear motivation for adopting either $h = 0.5$ 
or $n=1$ as standard values, this is the most common strategy in the 
literature.

\subsection{Single-parameter variations}

\subsubsection{Scale-invariant CDM models}

Since it was recognized that the CDM model could be fixed by lowering the 
shape parameter, considerable attention has been directed towards achieving 
this end by lowering the density parameter $\Omega_0$, usually retaining 
spatial flatness via the introduction of a cosmological constant. The 
alternative strategy to achieve this is to lower the Hubble parameter. With 
a slightly different motivation, such a strategy has been long advocated by 
Shanks (e.g. 1985). It was mentioned by Liddle \& Lyth \shortcite{LL93a} 
and proposed as a possibility more vigorously by Bartlett et al. 
\shortcite{BBST}. However, neither of those papers took advantage of 
the effect of the baryon content in these models, which can play a 
significant role in reducing the shape parameter, as given by equation 
(\ref{shape}). Consequently, it seems that the proposed $h \simeq 0.3$ may 
be too strict and we find that one can get away with $h \la 0.35$. This is 
still a long way from the values currently discussed via direct observation 
\cite{H0,H02}. However, the preferred baryon density may prove yet higher 
than the value we have adopted, say at the top of or beyond the range given 
by recent analyses of nucleosynthesis (Copi et al. 1995a,b), which would 
help to alleviate worries about the high baryon abundance in clusters if 
$\Omega_0$ does turn out to be 1 \cite{WNEF,WF}. Then high baryon density 
may become an increasingly attractive solution to the problems of standard 
CDM.

We remind the reader in passing that altering the number of massless species 
provides a way of mimicking the low-$h$ power spectrum \cite{DGT,WGS} while 
retaining a higher true value of $h$.

\subsubsection{Scale-invariant CHDM models}

The idea of introducing a hot dark matter component to reduce the 
short-scale power relative to CDM has a long history 
\cite{ShaS,BV,Fang,VB,H89,SSS,vDS} and it quickly received a lot of 
attention (Schaefer \& Shafi 1992, 1993; Davis, Summers \& Schlegel 1992; 
Taylor \& Rowan-Robinson 1992; Holtzman \& Primack 1993) after the {\it 
COBE} observations. The most widely explored versions of the CHDM model 
assume both $n=1$ and $h=0.5$. As far as detailed simulation is concerned, 
the bulk of attention has gone to the choice $\Omega_{\nu} = 0.3$ 
\cite{DSS,KHPR,JMBF,NKP,Yepes,Bryan,KNP}.

The alternative approach, as adopted in this paper, is to investigate the 
parameter space more widely by concentrating on linear theory, and this has 
been done in many recent papers (Taylor \& Rowan-Robinson 1992; Liddle \& 
Lyth 1993b; Pogosyan \& Starobinsky 1993, 1995a; Schaefer \& Shafi 1994). 
Our aim here is to examine the widest possible parameter space using the 
most up-to-date linear theory constraints.

Should the recently claimed detections of the muon neutrino mass be 
confirmed, then it corresponds to a particular region of our parameter 
space. Assuming the standard abundance calculation, the possible LSND 
detection \cite{Caldwell,PHKC} corresponds, with considerable uncertainty, 
to $\Omega_{\nu} = 0.1 (h/0.5)^{-2}$; if $h$ becomes smaller this 
corresponds to a greater fraction of the total density. 

A much advertized drawback of the $\Omega_{\nu} = 0.3$ CHDM model is the 
possibility that it may not reproduce the observed abundance of damped Lyman 
alpha systems \cite{MM,KC,MB}. This has led some to favour the reduction of 
$\Omega_{\nu}$ to 0.25 or 0.20 \cite{KBHP}.  We find that, were we to make 
the same assumptions as they do concerning the {\it COBE} normalization and 
the damped Lyman alpha system abundance, we would more or less reproduce the 
constraint of Klypin et al. \shortcite{KBHP}. Since their calculation is 
considerably more sophisticated than ours, being simulation based, one could 
regard this as a calibration of our calculation, though we have not had to 
do any tuning. Anyway, since their calculation was made, things have 
generally gone in the direction of weakening the early galaxy formation 
constraint on the CHDM model; the {\it COBE} normalization has gone up 
slightly, and more recent damped Lyman alpha system abundance observations 
\cite{storrie} have produced lower results than those of Lanzetta et al. 
\shortcite{LWT}. Consequently, we find that the constraint from damped Lyman 
alpha systems on {\it COBE}-normalized CHDM models has weakened somewhat, 
back to $\Omega_{\nu} \la 0.30$ for the $n=1$, $h=0.5$ version.

However, a more important question is what values of $\Omega_{\nu}$ are 
preferred when one brings other data into play. Fig.~4 shows that there are 
already indications from the shape of the galaxy correlation function that 
the $\Omega_{\nu} = 0.30$ model is subtracting too much short-scale power. A 
better eyeball fit to the correlation function data from Fig.~4 is 
$\Omega_{\nu} = 0.20$. This sort of value has been criticized on alternative 
grounds, that it may overproduce clusters \cite{PHKC}, a problem made worse 
by the higher normalization but slightly improved by our view that the 
cluster constraint is weaker than usually advertized. Fig.~5 illustrates our 
allowed parameter region as a function of both $\Omega_{\nu}$ and $h$. The 
combination of cluster and damped Lyman alpha system abundance appears 
sufficient to exclude all models with $h \geq 0.55$. For values of 
$\Omega_\nu 
\la 0.3$ the cluster constraint alone is what limits the value of $h$. [The 
model predictions of the cluster amplitude would, however, be compatible 
with the constraint for higher $h$ if $\Omega_\nu$ were composed of more 
than one degenerate mass flavour \cite{PHKC,BSS}.]  

Fig.~5 makes it clear that there is also a lot of extra freedom to 
be gained via fairly modest decreases in $h$, with a wide band of allowed 
values opening up. If the LSND detection corresponds to a single neutrino, 
giving $\Omega_{\nu} \sim 0.1 (h/0.5)^{-2}$, it cuts across the allowed 
region around $h=0.4$, with considerable uncertainty. 

Overall, our analysis suggests that, largely due to the higher {\it COBE} 
normalization, the parameter space of scale-invariant CHDM is not too 
large. However, we shall see that, when one allows $n$ to vary and includes 
the possibility of gravitational waves, the freedom becomes greater.

\subsection{Tilted CDM models}

Let us now extend the discussion to take in the general class of 
inflation-based cold dark matter models. Naturally, one chooses 
$\Omega_{\nu}$ to be zero, but the parameters $n$ and $h$ are to be freely 
varied and gravitational waves added if desired.

In the context of an arbitrary choice for the initial perturbation spectrum, 
the possibility of choosing a spectral index other than $n=1$, which has now 
become known as `tilt', has often been discussed. The modern context, where 
the origin of the tilt is identified as inflation and a connection made to 
the desired slope of the galaxy correlation function, was discussed by Bond 
\shortcite{Bond}, Liddle, Lyth \& Sutherland \shortcite{LLS}, Cen et al. 
\shortcite{CGKO}, Adams et al. \shortcite{ABFFO} and Liddle \& Lyth 
\shortcite{LL93a}. After the original {\it COBE} result came out \cite{COBE} 
the prognosis for such models was not particularly good: the necessary tilt 
to explain the galaxy correlation function, especially as witnessed by the 
APM survey \cite{APM}, left a perceived deficit of short-scale power when 
adjusted to the {\it COBE} data. Since then the situation has improved 
somewhat, due to the higher normalization of the current {\it COBE} data 
\cite{G96}. It appears from Fig.~4 that, for $h = 0.5$, even tilting 
to $n = 0.7$, a commonly discussed number, is not sufficient to get the 
slope of the galaxy correlation function right, and one has to go even 
lower. Substantial gravitational waves, as there would be in a power-law 
inflation model, would make things yet worse, so for values of $h$ near 0.5 
the implementation must be in a model such as natural inflation \cite{ABFFO} 
which predicts negligible gravitational waves.

Fig.~6 shows contour plots of the constraining observations in the 
$n$--$h$ plane. The top panel is with no gravitational waves; the lower 
panel adopts the power-law inflation amplitude of gravitational waves for 
$n<1$ and zero otherwise, as discussed in Section 2.3. This figure confirms 
the viability of scale-invariant CDM provided that $h$ is low enough. More 
importantly, it shows that there may still be a reasonable amount of 
parameter space available for CDM models. Provided that $n$ is lowered 
sufficiently, these can work for values of $h$ up to about 0.5, but not for 
any higher values. The parameter space widens out to low values of $h$, so 
no useful lower limit on $h$ can be obtained this way. The incorporation of 
gravitational waves reduces the favoured area.

Although we have fixed the baryon density, in the regime where the spectrum 
is well described via the shape parameter defined by equation (\ref{shape}) 
one can account for a variation by defining an effective $h$ value. For the 
range of $\Omega_{{\rm B}}$ allowed by nucleosynthesis the change is always 
small, and via a small parameter expansion one can write $h_{{\rm eff}} 
\simeq h(1-2\Delta\Omega_{{\rm B}})$, where $\Delta \Omega_{{\rm B}} = 
\Omega_{{\rm B}} - \Omega_{{\rm B}}^{{\rm nuc}}$. This allows one to 
interpret a point in Fig.~5 as representing a range of models with slight 
simultaneous variation in $\Omega_{{\rm B}}$ and $h$ satisfying this 
relation. However, within the nucleosynthesis range only small changes can 
be made.

White et al. \shortcite{WSSD} analysed two particular versions of the CDM 
model. In their favoured models, the desired reduction in short-scale power 
is brought about by accumulating small changes from all sources. They 
considered a slightly higher baryon density which can be incorporated as 
just discussed. Without gravitational waves, they favoured $n = 0.8$ and $h 
= 0.45$ (their higher $\Omega_{{\rm B}}$ giving $h_{{\rm eff}} = 0.43$), 
which is at the edge of our favoured parameter range. However, with 
gravitational waves they preferred $n = 0.9$, still with $h = 0.45$, which 
our results do not favour; our treatment of the shape of the galaxy 
correlation function is more stringent than theirs and a smaller value of 
$h$ would be required.

\subsection{Tilted CHDM models}

The full parameter space is best investigated by running slices through the 
volume of $n$--$h$--$\Omega_{\nu}$ space. Following Pogosyan \& Starobinsky 
\shortcite{PogStar2}, we show cuts of constant $n$ and of constant $h$. We 
do each of these for both the case without gravitational waves and the case 
with power-law inflation gravitational waves.

Figs 7 and 8 show the slicings for the case of no gravitational waves. 
Fig.~7 includes a reproduction of the scale-invariant CHDM case shown in 
Fig.~5. As anticipated from all the special cases we have already examined, 
there is a fairly reasonable parameter space available which explains all 
the observational data. Unless $h$ is below 0.5, a component of HDM appears 
to be required. Larger values of $h$, up to around 0.65 in the most extreme 
cases, are then permitted provided that one introduces a strong tilt as well 
the hot component.

As regards $\Omega_{\nu}$, it seems that the highest it can reach is $0.35$ 
in the rather extreme case of $h=0.4$ and $n \simeq 1.2$. For $n$, the 
lowest working value is $n=0.6$, while concerning high values $n$ above 
$1.2$ is possible provided that a very low value of $h$ is tolerated. In 
this regard, our conclusions favour those of Pogosyan \& Starobinsky 
\shortcite{PogStar2} rather than Lucchin et al. \shortcite{Letal} in that we 
see no particular advantage in adopting a blue ($n>1$) spectrum and find 
that values above $1.1$ can be maintained only via a dubiously low Hubble 
parameter. Borgani et al. \shortcite{BLMM} have suggested that the adoption 
of a blue spectrum helps to alleviate worries about damped Lyman alpha 
system abundance; while in isolation this is certainly true we find that 
this strategy is not favoured by other data\footnote{After our paper was 
submitted, Borgani et al. \shortcite{BLMM} was revised to include discussion 
of the cluster abundance, obtaining conclusions similar to ours.}.

Figs 9 and 10 show equivalent slicings for the case where power-law 
inflation gravitational waves are included for $n < 1$. For $n \geq 1$ they 
are the same as Figs.~7 and 8. They show that $n$ has to be above $0.8$ 
before any sort of fit to the data is available. This is part of a fairly 
general result that the incorporation of gravitational waves reduces the 
total available parameter space. Only a few narrow slivers of extra 
parameter space are opened up by the inclusion of gravitational waves. This 
is because of the strength of the shape parameter constraint, which forces a 
fairly dramatic reduction in short-scale power. Models able to fit this are 
typically not able to lose much more of this power by permitting some of the 
{\it COBE} signal to be soaked up by gravitational waves.

\section{Discussion}
\label{DISC}

Despite the recent attention directed at low-density models of structure 
formation, either open or with a cosmological constant, the possibility of 
working models with the critical density remains an attractive one. We have 
been able to show that the current observational constraints continue to 
allow a substantial amount of parameter space for these models.

An important sub-class of the models that we have discussed is generalized 
CDM models. Here all the dark matter is assumed to be cold, and one attempts 
to fit the data by permitting variation of the Hubble parameter and the 
initial conditions (tilt and gravitational waves) coming from inflation. We 
illustrated the constraints in Fig.~6; they demonstrate that there is still 
an area of parameter space available for CDM models. The most plausible 
situation requires a tilt to $n<1$ and not too high a gravitational 
wave amplitude. The biggest drawback these models face is that they require 
a low value of the Hubble parameter; a strong tilt just permits $h = 0.5$, 
but that is the highest in any region of parameter space. The bulk of 
available parameter space, where the tilt is not so strong, requires $h$ 
some way below this. This sits uncomfortably with recent direct measurements 
of the Hubble constant \cite{H0,H02}, though we remind the reader again that 
power spectra mimicking low values of $h$ can be achieved by introducing 
extra massless species/decaying particles. 

In the case of the scale-invariant CHDM models, we find that they remain 
viable at $h = 0.5$, but hardly any higher. However, when one 
introduces the additional freedom of varying $n$, a much more substantial 
parameter region opens up which does allow fits at higher values of $h$. A 
value $h = 0.6$, consistent with recent direct measurements, seems easy to 
achieve provided that one introduces a strong enough tilt along with the HDM 
component, and perhaps even $h=0.65$ is possible if one pushes right to
the corner of the parameter space (though the age of the universe in such a 
model could be problematic).

We have not investigated the introduction of gravitational waves as 
thoroughly as the other parameters, but we have looked at the case where the 
amplitude is that given by power-law inflation. What we find is that the 
gravitational waves are not very helpful; they lead to a reduction in the 
allowed parameter `volume' and only in very limited regions do they allow 
working models for values of $n$, $h$ and $\Omega_{\nu}$ that would fail 
without gravitational waves. However, that said, even with this rather large 
gravitational wave component there are still significant allowed regions. 
Large-scale structure is therefore not able to exclude the possibility of 
such gravitational waves.

To conclude, we have presented an extensive comparison of critical-density 
models for structure formation with linear and quasi-linear observational 
data. We have calculated new transfer functions and provided an improved 
fitting formula for them which gives the power spectra as a continuous 
function of all of $n$, $h$, $\Omega_{\nu}$, $\Omega_{{\rm B}}$ and $z$. We 
have interpreted the data in terms of the dispersion of the density contrast 
smoothed on some scale $R$. We have found a substantial allowed 
parameter space for CDM and CHDM models, which at least in the latter case 
seems likely to survive for some time to come. Critical-density models 
continue to offer a viable and aesthetically simple basis for understanding 
structure formation.

\section*{Acknowledgments}

ARL is supported by the Royal Society, RKS and QS by DOE DE-FG02-91ER, 
NASA NAG5-2646 and the Bartol Research Institute, and PTPV by the PRAXIS XXI 
programme of JNICT (Portugal). ARL and DHL thank the Aspen Center for 
Physics, where discussions pertinent to this paper took place. We thank 
Stefano Borgani, Martin Hendry, Lev Kofman, Cedric Lacey, John Peacock, 
Dmitry Pogosyan, Joel Primack, Douglas Scott and Martin White for helpful 
discussions on a variety of topics. ARL and PTPV acknowledge the use of the 
Starlink computer system at the University of Sussex.


\bsp
\section*{Figure Captions}

\vspace*{12pt}
\noindent
{\bf Figure 1:} We plot the dispersion $\sigma(R,z)$ for two different 
CHDM models, $\Omega_{\nu} = 0.2$ and $0.3$, at redshifts of zero (solid 
lines) and 3.5 (dashed lines). The lower curves correspond to $\Omega_{\nu} 
= 0.3$. The curves have been normalized on to each other at large scales, 
which is achieved by using the CDM growth law $\sigma(R,z) \propto 
(1+z)^{-1}$. Except for the shortest scales, the transfer functions are 
redshift-independent, indicating that the CDM growth law holds for $R \geq 3 
h^{-1}$ Mpc.

\vspace*{12pt}
\noindent
{\bf Figure 2:} The observational data that we consider, interpreted in 
terms of $\sigma(R)$. Error bars are 1$\sigma$ and lower limits are 95 per 
cent confidence. We represent the {\it COBE} data schematically at $4000 
h^{-1}$ Mpc as discussed in text; they are indicated by a filled square the 
size of which roughly represents the uncertainty. The Peacock \& Dodds data 
are shown by circles (as discussed, we omit the leftmost four points); there 
is an uncertainty in overall normalization which has not been illustrated. 
The bulk flow constraint is represented by a star, and the cluster abundance 
constraint by a cross. The lower limits on the left hand side correspond to 
damped Lyman alpha systems (leftmost, values for redshifts 3 and 4 overlap) 
and quasars (right). Although the data clearly show a smooth trend, they 
cover such a range in $\sigma(R)$ values that one cannot use a figure of 
this form to compare models by eye.

\vspace*{12pt}
\noindent
{\bf Figure 3:} As Fig.~2, but plotting $\sigma(R)$ relative to its value 
in the {\it COBE}-normalized standard CDM model. This greatly improves 
clarity. The data points are as in Fig.~2 [{\it COBE}, filled square; bulk 
flows, star; cluster abundance, cross; damped Lyman alpha system (now shown 
at two different redshifts) and quasar abundances, lower limits] except that 
we now show the Peacock \& Dodds data as a band representing the 1$\sigma$ 
errors about the (unplotted) central values. The error bars on the end of 
the band indicate their estimate of the uncertainty in overall normalization 
of this data set. We see that the data are not well fitted by the standard 
CDM model, which possesses too much short-scale power. Although to a 
reasonable accuracy this figure applies at any epoch, it is most accurately 
applied at redshift 3.5 corresponding to the quasar and damped Lyman alpha 
system abundances, so that one need not worry about the suppressed 
perturbation growth rate in models with an HDM component.

\vspace*{12pt}
\noindent
{\bf Figure 4:} The data plotted as in Fig.~3, with some illustrative 
theoretical curves overlaid for comparison. These curves are those 
appropriate to redshift 3.5, as discussed in the text. We have shown only
examples where a single parameter of the standard CDM scenario has been 
modified. The solid line is the standard CDM model; the others modify one 
parameter from this fiducial model, as indicated in the key. All models are 
precisely {\it COBE} normalized; the {\it COBE} point at $4000h^{-1}$ Mpc is 
illustrative. Remembering that one can shift the entire Peacock \& Dodds 
data set vertically, reasonable eyeball fits to the data are possible via 
any of the following: lowering $h$ to about 0.35, lowering $n$ to about 0.7 
assuming no gravitational waves, introducing a hot dark matter component at 
about the $\Omega_{\nu} = 0.2$ level.

\vspace*{12pt}
\noindent
{\bf Figure 5:} Scale-invariant CHDM models. The lines shown are from the 
galaxy correlation data (dotted), cluster abundance (dashed) and damped 
Lyman alpha systems (solid). Shading indicates the favoured area. All 
constraints are plotted at 95 per cent confidence.

\vspace*{12pt}
\noindent
{\bf Figure 6:} Contour plots of constraining observations for CDM models. 
The upper panel is without gravitational waves; the lower panel includes 
power-law inflation gravitational waves for $n < 1$. The lines shown are 
galaxy correlations (dotted), cluster abundance (dashed), damped Lyman alpha 
system abundance (solid) and POTENT (dot-dashed). Shading indicates the 
favoured area and all data are plotted at 95 per cent confidence.

\vspace*{12pt}
\noindent
{\bf Figure 7:} Four slices through the $h$--$\Omega_{\nu}$ plane at 
different $n$, with no gravitational waves included. The values of $n$ are 
as indicated, and the data are plotted as in Fig.~6.

\vspace*{12pt}
\noindent
{\bf Figure 8:} Four slices through the $n$--$\Omega_{\nu}$ plane at 
different $h$, with no gravitational waves included. The values of $h$ are 
as indicated, and the data are plotted as in Fig.~6.

\vspace*{12pt}
\noindent
{\bf Figure 9:} Four slices through the $h$--$\Omega_{\nu}$ plane at 
different $n$, with gravitational waves included. The values of $n$ are as 
indicated, and the data are plotted as in Fig.~6.

\vspace*{12pt}
\noindent
{\bf Figure 10:} Four slices through the $n$--$\Omega_{\nu}$ plane at 
different $h$, with gravitational waves included. The values of $h$ are as 
indicated, and the data are plotted as in Fig.~6.

\end{document}